\documentclass[%
 aip,
 amsmath,amssymb,
 reprint,%
]{revtex4-1}

\usepackage[pdftex]{graphicx,color}
\usepackage{dcolumn}
\usepackage{bm}

\usepackage[utf8]{inputenc}
\usepackage[T1]{fontenc}
\usepackage{mathptmx}
\usepackage{etoolbox}
\usepackage{siunitx}
\usepackage{color}
\usepackage{ulem}

\makeatletter
\def\@email#1#2{%
 \endgroup
 \patchcmd{\titleblock@produce}
  {\frontmatter@RRAPformat}
  {\frontmatter@RRAPformat{\produce@RRAP{*#1\href{mailto:#2}{#2}}}\frontmatter@RRAPformat}
  {}{}
}%
\makeatother
\begin{document}

\preprint{AIP/123-QED}

\title[]{Disordering two-dimensional magnet-particle configurations using bidispersity}
\author{K. Tsuchikusa}
 \affiliation{Department of Earth and Space Science, Osaka University, 1-1 Machikaneyama, Toyonaka 560-0043, Japan}

\author{K. Yamamoto}%
 \affiliation{Department of Earth and Space Science, Osaka University, 1-1 Machikaneyama, Toyonaka 560-0043, Japan}
 \affiliation{Water Frontier Research Center (WaTUS), Tokyo University of Science, 6-3-1 Niijuku, Katsushika-ku, Tokyo 125-8585, Japan}

\author{M. Katsura}
 \affiliation{Department of Earth and Space Science, Osaka University, 1-1 Machikaneyama, Toyonaka 560-0043, Japan}

\author{C.T. de Paula}
 \affiliation{Departamento de Matem\'atica, Universidade de Brasília, Campus Universit\'ario Darcy Ribeiro, Brasília, DF 70910-900, Brazil}

\author{J.A.C. Modesto}
 \affiliation{Departamento de Matem\'atica, Universidade de Brasília, Campus Universit\'ario Darcy Ribeiro, Brasília, DF 70910-900, Brazil}

\author{S. Dorbolo}
 \affiliation{GRASP, Institute of Physics, Building B5a, Sart Tilman, Université de Li\`ege, B4000 Li\`ege, Belgium}

\author{F. Pacheco-V\'azquez}
 \affiliation{Instituto de F\'isica, Benem\'erita Universidad Aut\'onoma de Puebla, Apartado Postal J-48, 72570 Puebla, Mexico}

\author{Y.D. Sobral}
 \affiliation{Departamento de Matem\'atica, Universidade de Brasília, Campus Universit\'ario Darcy Ribeiro, Brasília, DF 70910-900, Brazil}

\author{H. Katsuragi$^*$}
\email{katsuragi@ess.sci.osaka-u.ac.jp}
 \affiliation{Department of Earth and Space Science, Osaka University, 1-1 Machikaneyama, Toyonaka 560-0043, Japan}

\date{\today}

\begin{abstract}
In various types of many-particle systems, bidispersity is frequently used to avoid spontaneous ordering in particle configuration. 
In this study, the relation between bidispersity and disorder degree of particle configuration is investigated. 
By using magnetic dipole-dipole interaction, magnet particles are dispersed in a two-dimensional cell without physical contact between them. 
In this magnetic system, bidispersity is introduced by mixing large and small magnets. 
Then, the particle system is compressed to produce a uniform particle configuration. The compressed particle configuration is analyzed by using Voronoi tessellation for evaluating the disorder degree which strongly depends on bidispersity. 
Specifically, standard deviation and skewness of the Voronoi cell area distribution are measured. 
As a result, we find that the peak of standard deviation is observed when the numbers of large and small particles are almost identical.
Although the skewness shows non-monotonic behavior, zero skewness state (symmetric distribution) can be achieved when the numbers of large and small particles are identical. 
In this ideally random (disordered) state, the ratio between pentagonal, hexagonal, and heptagonal Voronoi cells become  roughly identical, while hexagons are dominant in monodisperse (ordered) condition.
The relation between Voronoi cell analysis and the global bond orientational order parameter is also discussed.
\end{abstract}

\maketitle

\section{\label{sec:level1}Introduction}

When modeling behaviors of many-particle systems, monodisperse system usually results in an ordered structure. 
However, most of the natural particle systems exhibit a random particle configuration, in general.
To achieve a random particle configuration, two types of particles (usually large and small) are mixed as a bidispersive system.
In a previous study~\cite{forcechain}, relation between bidispersity and particle configuration was investigated through the ordering of force chain orientations using the photoelastic effect which allows a direct access to the force chain network. 
According to their result, the order parameter of the force-chain orientational order becomes roughly constant within the range of 10-90\% bidispersity (defined as the ratio of area occupied by small (large) disks to total area of the disks). This result suggests that even a 10\% bidispersity is adequate to generate random particle configurations. It is important to note, however, that the randomness of the force chain structure and particle configuration may not coincide. The preceding investigation solely examined the force-chain orientational order.  
To properly characterize the particle configuration order in detail, we should directly measure and analyze the particle configuration as well.

Bidisperse systems have been used in various granular studies. 
Usually, bidispersity is advantageous to prevent ordering with crystalline configuration.
Moreover, other structures such as quasicrystals and DNA-like structures can also be developed in bidisperse systems~\cite{Reinhardt:2017,Fayen:2020,Fayen:2022}. 
To understand such complex structure formation processes, the effect of bidispersity on the disordering of particle configuration has to be revealed. 
In this study, we are going to focus on the magnet particle system, in which ordered configuration formed by monodisperse system can be disturbed by introducing bidispersity. By using binary mixture of two types of dipolar particles, a diverse array of crystalline or partially ordered structures can be generated~\cite{Assoud:2007,Messina:2016,Ebert:2008,Fornleitner:2008,Fornleitner:2009,Schockmel:2019}. Under certain conditions, partial clustering~\cite{Ebert:2009} or relaxation after ultrafast quenching~\cite{Assoud:2009} can also be observed. While the characterization of various types of ordered structures is an interesting research topic, the current study aims to focus on the characterization of random structures through analyzing distribution of particle configuration.

The evaluation of the ordered configuration has been extensively studied in terms of melting (solid-liquid) transition~\cite{Strandburg:1988,Gasser:2010,Schockmel:2013,Messina:2015,Schockmel:2017,Opsomer:2020,Zahn:1998,Gribova:2011,Komatsu:2015}. 
Particularly, in two-dimensional melting systems, completely ordered (crystalline) structure is disrupted by the dislocation pairs of defects. 
Then, the completely disordered (random) structure is attained in the molten state. 
The hexatic state can be observed in between these ordered and disordered states. 
Various melting-related phenomena have been studied by using the concept of two-dimensional melting model~(e.g.~\cite{Strandburg:1988,Gasser:2010}). 
A more or less similar two-step melting-like behavior was found in a vibrated granular 2D layer~\cite{Reis:2006}. 
To characterize the hexatic phase, the bond orientational order correlation function has been used~\cite{Gasser:2010}. 
In the molten state, hexagonal order structure does not have long range order. 
Thus, the bond orientational correlation function shows exponential decay in space. 
Contrastively, the bond orientational correlation function becomes constant in the solid crystalline state. 
The bond orientational correlation function shows power-law decay in hexatic states. 
Similarly, the radial distribution function has also been used to characterize the translational ordering degree of particle configuration. 

Recently, solidification of the system with or without ordering has been studied also in terms of jamming transition~\cite{Behringer:2018}. 
Even biological systems show the jamming-like behavior~\cite{Bi:2016}. 
In the context of jamming transition, physical mechanism of solidification with disordered structure has to be properly understood.
Thus, a method to characterize the ordering/disordering in the jammed system should be developed. 
Particularly, when the number of particles is limited, simpler and more efficient ways of characterizing the order of particle configuration are necessary. 
Although bidisperse systems have been used to mimic natural polydisperse systems, very weak bidispersity may not be adequate to attain a spatially homogeneously random state. Such weak bidispersity may cause spatially heterogeneous randomness, which could affect the physical behaviors of jammed systems composed of bidisperse particles. The current study aims to investigate the relation between the degree of bidispersity and the homogeneity of the random structure. 

Granular systems have been well studied as a typical example of macroscopic many particle systems. 
In usual granular systems, friction due to particle-particle contact plays a major role to construct and keep the particle configuration.
Recently, different types of granular materials consisting of magnetically repulsive particles have also been used to study non-contacting granular behaviors~\cite{magnet,Discharge,viscoelastic}.
For instance, two-dimensional magnet particle systems have been utilized to study granular silo flow~\cite{magnet, Discharge}, impact drag force~\cite{viscoelastic}, and slow compression mechanics~\cite{compression}.
Among magnetic particles, physical contact can be suppressed and interactions can be mediated only by repulsive magnetic forces.
Therefore, the particle configurations can be easily rearranged. In addition, this non-contacting feature makes the measurement and analysis of particles easier.
In the present study, therefore, we investigate the effect of bidispersity in a two-dimensional magnetic particle system in which there is no inter-particle contact.

In short, we study a two-dimensional system consisting of magnet particles that have repulsive interaction. 
This system can develop various ordered and disordered particle configurations by varying the degree of bidispersity. 
In this study, the number of particles is limited and the aspect ratio of the system is large. 
Therefore, we characterize the particle configuration by a simple method. 
A series of experiments are conducted, and a comprehensive analysis of the particle configuration is performed using Voronoi tessellation and global bond orientational order. Through the use of Voronoi cell area distributions, this study demonstrates that the moments of the distribution are effective and straightforward tools for characterizing the spatial homogeneity of disordered structures in the granular packing. 

\section{Experiments}

\begin{figure}
\includegraphics[width = 85mm]{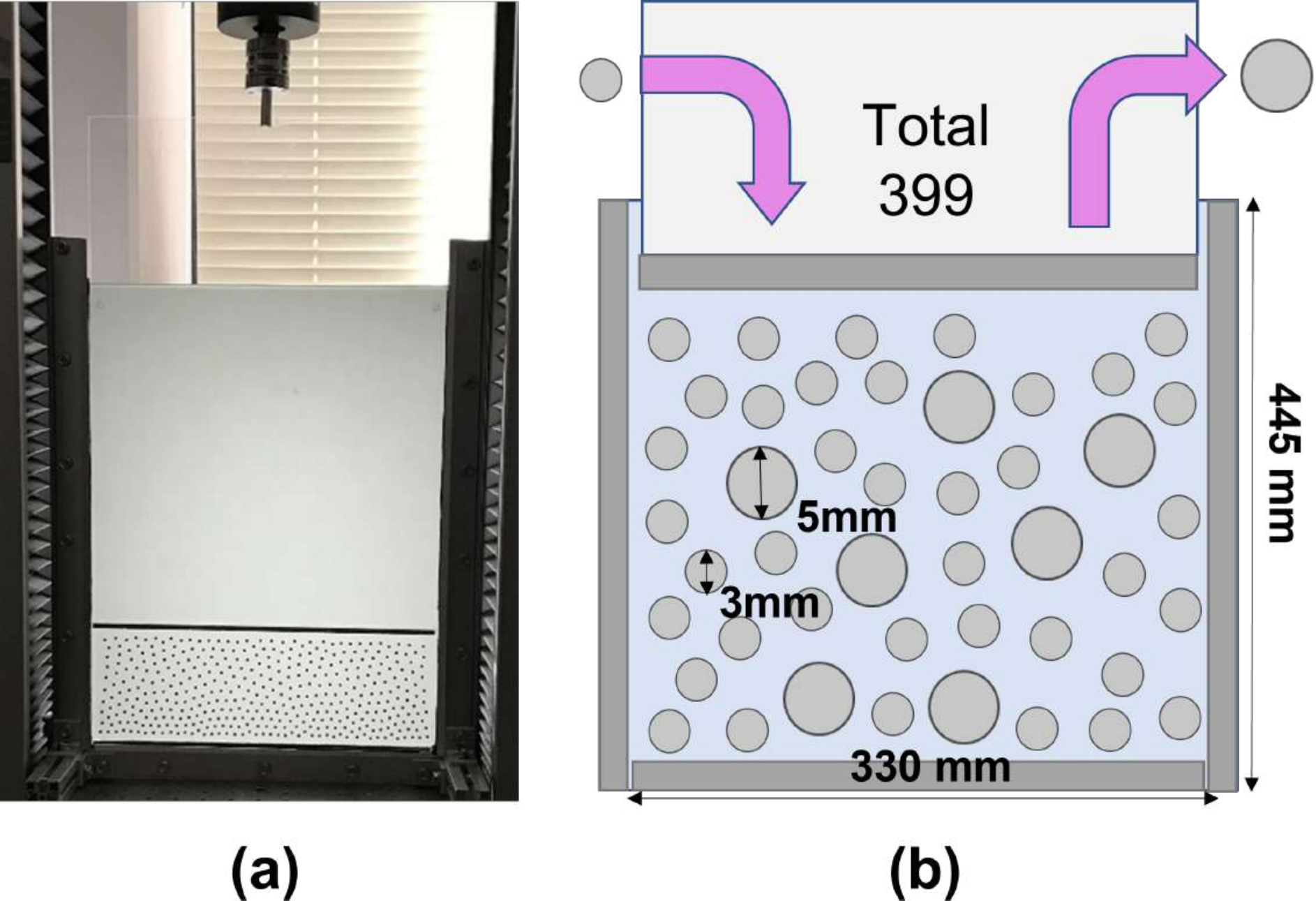}
\caption{\label{fig:setup} Photograph and schematic diagram of the experimental equipment: (a) front view image and (b) schematic diagram with dimensions of the cell and magnet particles.}
\end{figure}

\begin{figure}
\includegraphics[width = 85mm]{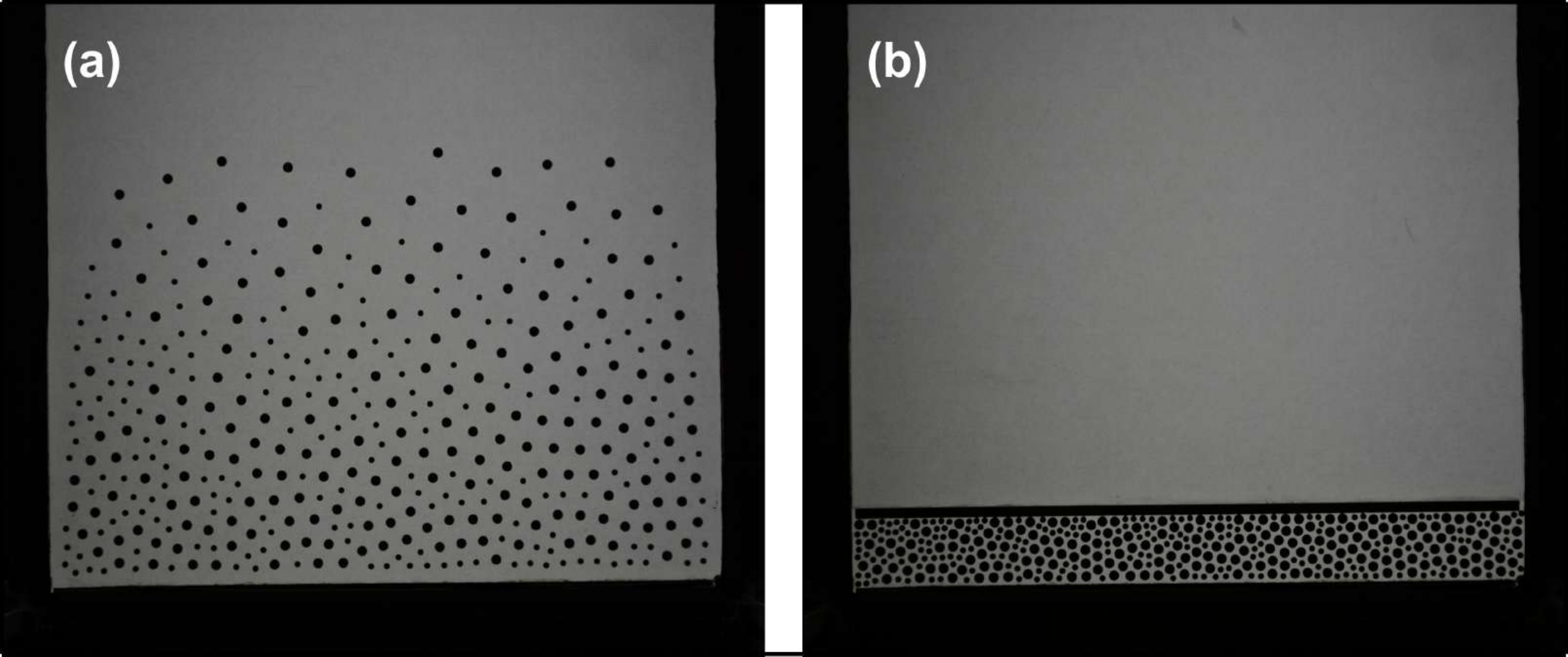}
\caption{\label{fig:raw_data} Raw data of the experiment: 199 Large particles and 200 small particles are distributed in the cell ($R=0.5$). Photographs of (a)~before compression and (b)~after compression are shown.}
\end{figure}

The experimental apparatus containing magnetic particles is a Hele-Shaw cell with transparent glass walls.
FIG.~\ref{fig:setup} shows the experimental apparatus: (a) front view and (b) schematic diagram.
Two glass walls of \SI{450}{mm} high, \SI{340}{mm} wide, and \SI{8}{mm} thick are used to construct the cell. 
Two glass plates are kept parallel with a gap width of \SI{3,0}-\SI{3,1}{mm}.
The cell is vertically mounted on a universal testing machine (Shimadzu AGX).

Then, cylindrical magnets are dispersed into the cell.
We use two types of NdFeB magnet particles.
One is \SI{5}{mm} in diameter, \SI{3}{mm} in thickness, and surface magnet flux density of \SI{4000}{G} (hereafter referred to as "large particles"). The other is \SI{3}{mm} in diameter, \SI{3}{mm} in thickness, and surface  magnet flux density of \SI{4100}{G} (hereafter referred to as "small particles").
All of these magnet particles are carefully placed so that the magnetic dipoles are perpendicular to the front and rear glass walls with identical direction of the magnetic moment.
Therefore, the magnet particles repel each other, without any physical contact. 
Rectangular NdFeB magnets are glued on the bottom and side walls to create repulsive boundary condition.
This configuration restricts the particle motions within a two-dimensional space.
The magnetic particles are manually distributed in the cell to form random initial structure.
To effectively fill the cell with a large number of magnet particles as much as possible, the magnet-particle configurations are manually compressed.
Then, the particles initial configuration is determined by the repulsive force balance between particles.
An acrylic plate (2~\si{mm} in thickness) is used as a piston to compress the particle system.
Rectangular NdFeB magnets are glued on the bottom of the piston. 
The experimental system used in this investigation is akin to the system used in Ref.~\cite{compression}, which primarily examined the behavior of the compression force. However, this study concentrates on characterizing the order/disorder structure of particles within the compressed system. 

In all experiments, the total number of particles is kept at $399$.
First, the experiment is performed with $399$ large particles. 
It is difficult to add more particles without external loading due to limitations in the system size. The dimensions of the cell are limited by the frame dimensions of the universal testing machine we use in this study. 
We prepare the initial particle configurations with the above-mentioned procedure.
The particle bed is then compressed by the piston at a constant speed of \SI{1}{mm.s^{-1}}.
The compression is terminated when the compression force reaches \SI{110}{N}, and the images of particle configurations are taken from the front of the cell by a still camera (Nikon, D7200).
The acquired image size is 6,000$\times$4,000 pixels. 
Spatial resolution of the acquired images is about $74$--$99$~$\mu$m/pixel. 
After unloading the particle system, a fixed number of large particles is replaced by an identical number of small particles. 
By this protocol, the ratio of the number of large particles and small particles can be varied while keeping the total number of particles constant.
After replacing the particles, the magnets in the cell are stirred with a bamboo stick to make a random structure and to eliminate the memory of particle configuration in the previous compression. Then, the system is compressed and the particle configuration is measured.

The aforementioned procedure is repeated until all particles are replaced by small particles.
To check the reproducibility, two or more experimental runs are performed for each experimental condition. 
In the following data plots, all results of these experimental runs are shown in scatter plots. 
The mixing ratio $R$ is defined by the ratio,
\begin{equation}
R= \frac{N_\mathrm{small}}{N_\mathrm{total}},
\label{eq:R_def}
\end{equation}
where $N_\mathrm{small}$ and $N_\mathrm{total}=399$ are the number of small particles and the total number of particles, respectively.
The ImageJ software is used to locate the position of these particles.
FIG.~\ref{fig:raw_data} shows the example images before and after compression for the state of $R=0.5$~(large 199 and small 200).

To stably compute the spatial correlation functions such as radial distribution function, the number of particles should be large and the system (boundary) should be more or less isotropic. 
However, in this study, we use a relatively small number of magnet particles confined in a thin (elongated) cell. 
Therefore, it is difficult to simply analyze the spatial correlation functions. Instead of spatial correlation functions, here we use a method characterizing global structure through the average of local characteristic quantities. The Voronoi tessellation and global bond orientational order (average of local bond orientational order) are more appropriate for extracting the geometrical characteristics of the assemblies~\cite{alloy,Finney:1977,Finney:1970}.

\begin{figure}
\includegraphics[width = 85mm]{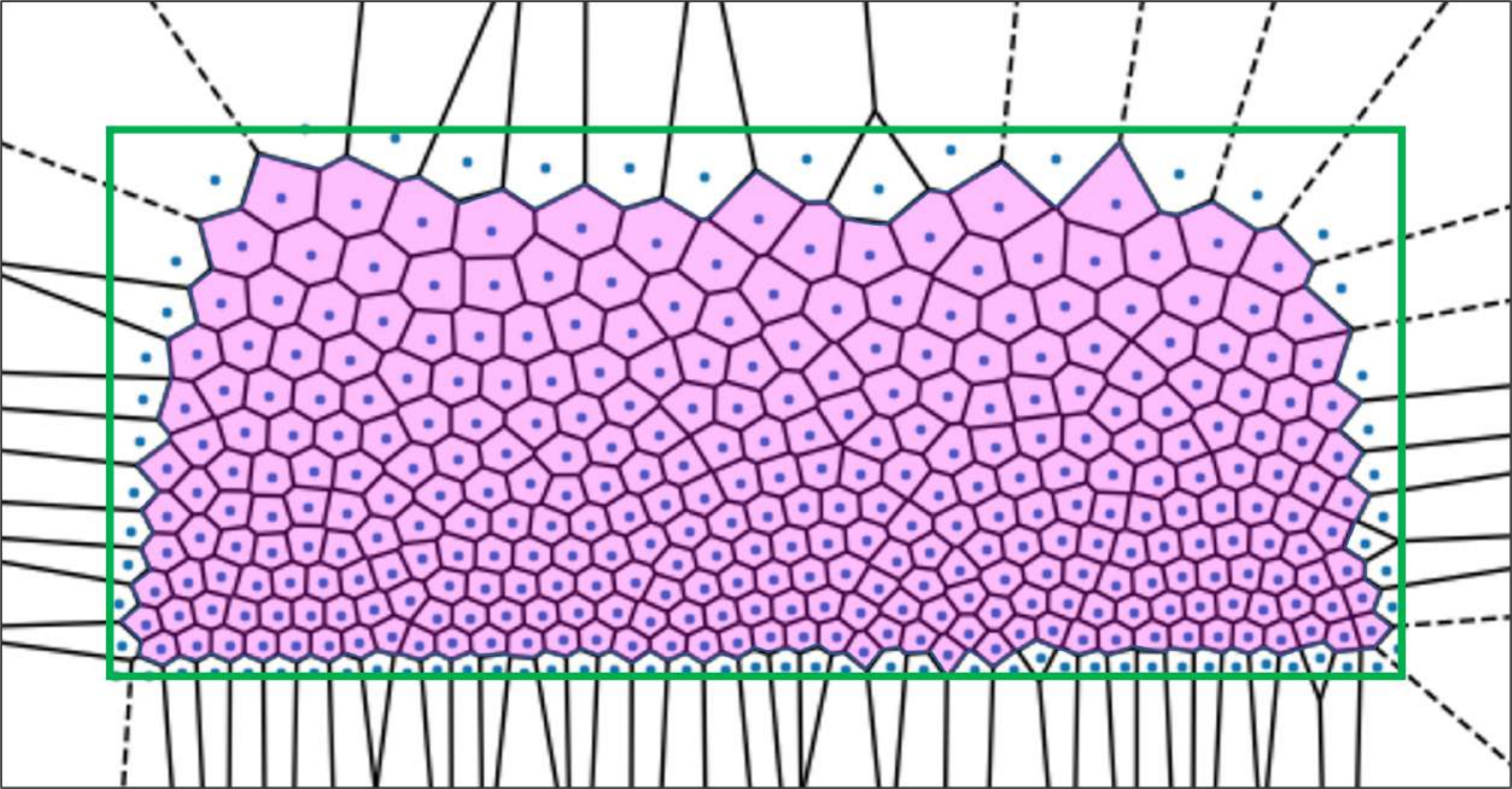}
\caption{\label{fig:Voronoi_example} An example of Voronoi tessellation based on the particle positions ($R$=0.5, before compression) The blue dots represent the center of the particle, the black lines represent the Voronoi partitions. Only filled (pink) Voronoi cells are analyzed.}
\end{figure}

An example of Voronoi tessellation is shown in FIG.~\ref{fig:Voronoi_example}~(pre-compression state of $R=0.5$). 
In this figure, the blue dots represent the positions of the center of the magnet particles.
The blue points are used as the reference generating points, and the plane coordinates are divided by line segments corresponding to the perpendicular bisectors of the neighbor generating points. 
In FIG.~\ref{fig:Voronoi_example}, the Voronoi cells near the walls diverge or have very large areas due to boundary effects.
Therefore, the green rectangle in FIG.~\ref{fig:Voronoi_example} that contains the topmost, bottommost, leftmost, and rightmost particles on the boundary is used to define the analysis area.
We consider only the Voronoi cells whose entire shapes are in this rectangle as shown by pink colored cells in FIG.~\ref{fig:Voronoi_example}.

Using the generating points, the global bond orientational order parameter of the configuration can also be defined as~\cite{Schockmel:2017},
\begin{equation}
  \langle \Psi_6 \rangle = \left\langle \left| \frac{1}{n}\sum_{j=1}^n e^{i6\theta_{ij}} \right| \right\rangle_i,
\label{eq:Psi_6_def}
\end{equation}
where $n$ represents the number of neighbors of $i$th particle, $\theta_{ij}$ is the angle between the horizontal axis and the bond linking $i$th particle and $j$th neighbor, and $\langle \cdot \rangle_i$ means the average of all particles. $\langle \Psi_6 \rangle$ becomes unity when the particle configuration obeys to a completely ordered triangular (hexagonal) lattice structure. Note that the value of $\langle \Psi_6 \rangle$ decreases upon the introduction of randomness or specific types of crystalline structure in the system.

\section{Results}

\begin{figure*}
\includegraphics[width = 180mm]{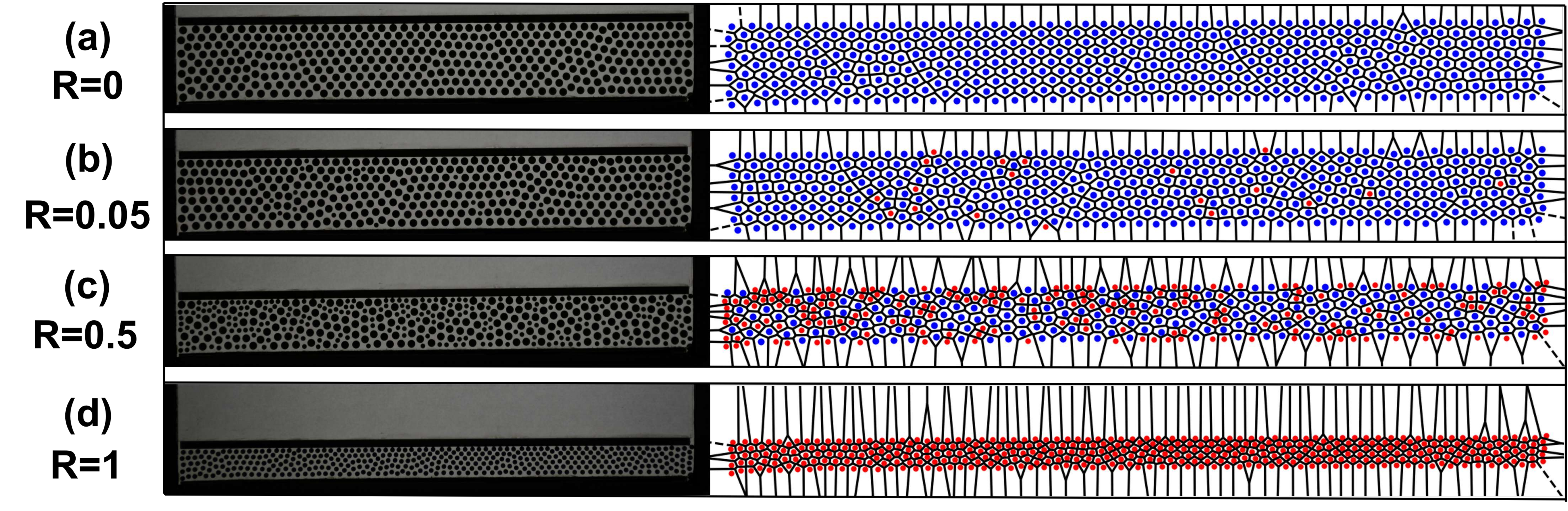}
  \caption{\label{fig:configs} Particle configurations of the compressed magnet particle systems with various mixing ratios. The left column shows the raw data, and the right column shows the corresponding Voronoi diagrams. Blue and red dots represent the positions of large and small particles, respectively. (a) $R=0$, (b) $R=0.05$, (c) $R=0.5$, and (d) $R=1$.
When $R=0$ and $1$, the height of compressed particle configuration comprises only about $9$ and $5$ rows of particles, respectively. Thus, many particles are affected by boundary walls.
  }
\end{figure*}

As shown in FIG.~\ref{fig:raw_data}(a) and \ref{fig:Voronoi_example}, particle configuration before the compression is not uniform due to the effect of gravity. Because we would like to focus on the effect of bidispersity, we analyze the configurations after the compression and discuss the dependence on $R$. 
Raw images of particle configurations for $R=0$, $R=0.05$, $R=0.5$, and $R=1$ of the compressed states (110~N compression) are shown in the left column of FIG.~\ref{fig:configs}.
When $R=0$ (FIG.~\ref{fig:configs}(a)), the particles are arranged in an ordered structure of hexagonal array.
At $R=0.05$ (FIG.~\ref{fig:configs}(b)), where there are a few small particles mixed in the system, the ordered arrangement of the particles is slightly disrupted compared to the state of $R=0$.
In addition, the ratio of non-hexagonal arrangement increases.
At $R=0.5$ (FIG.~\ref{fig:configs}(c)), where the numbers of large and small particles are almost identical, the particle configuration shows a disordered arrangement.
At $R=1$ (FIG.~\ref{fig:configs}(d)), where all particles are small, the particle configuration order is recovered, and the arrangement is again almost hexagonal. However, the degree of ordering is smaller than $R=0$ state because the small size particles have relatively greater individual variation in magnitude of the magnetic dipole moment. Individual variation of magnetic moment results in effective polydispersity.
Furthermore, since the small-particle system is significantly compressed and many particles are next to the walls, the boundary effect (including friction with the walls) is also enhanced. 
Specifically, the height of compressed particle configuration reduces to only $5$ layers of particles when $R=1$. Therefore, many particles are affected by boundary walls.

Voronoi diagrams for $R=0$, $R=0.05$, $0.5$, and $1$ are shown in the right column of FIG.~\ref{fig:configs}.
At $R=0$, the Voronoi cells are basically uniform in size. 
Besides, most of the Voronoi cells are hexagonal.
At $R=0.05$, some Voronoi cells are small.
In addition, the number of non-hexagonal Voronoi cells seems to increase.
At $R=0.5$, the Voronoi cells have various areas, and there are many non-hexagonal cells.
At $R=1$, the particles arrangement recovers the ordering. The number of hexagonal Voronoi cells also increases in this state.

\begin{figure}
\includegraphics[width = 90mm]{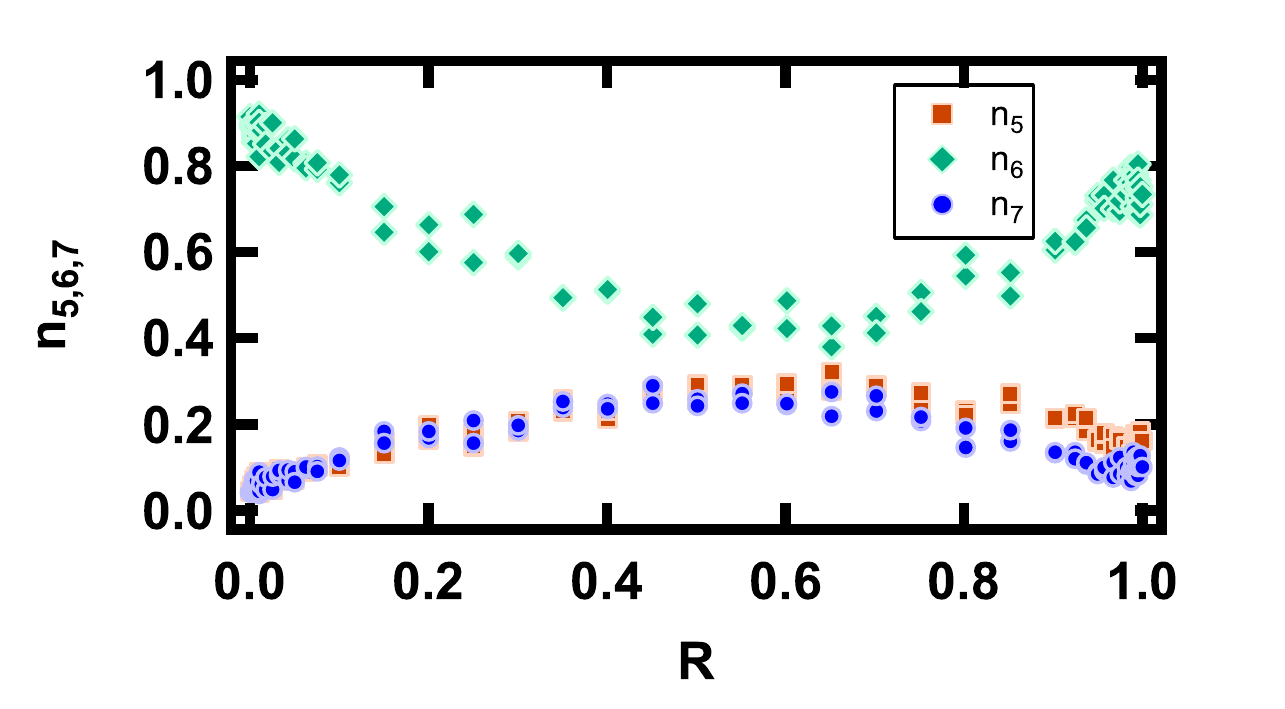}
\caption{\label{fig:poly_ratio} The mixing ratio dependence of the number ratios of pentagons, hexagons, and heptagons of Voronoi cells, normalized to the total number of pentagons, hexagons, and heptagons. Horizontal axis shows the mixing ratio $R$ and vertical axis shows the ratios of pentagons $n_5$, hexagons $n_6$, and heptagons $n_7$. }
\end{figure}

To characterize the observed features, we examine the fraction of hexagonal Voronoi cells with respect to the mixing ratio, $R$.
The number ratios of pentagonal, hexagonal, and heptagonal Voronoi cells to the total number of the Voronoi cells are defined as $n_5$, $n_6$, and $n_7$, respectively. There are few Voronoi cells other than pentagonal, hexagonal, and heptagonal cells.
FIG.~\ref{fig:poly_ratio} shows the $R$ dependence of $n_5$, $n_6$, and $n_7$.

As seen in FIG.~\ref{fig:poly_ratio}, $n_6$ is large when $R=0$, and decreases as $R$ increases. 
Then, $n_6$ reaches the minimum at $R\approx 0.5$.
After that, $n_6$ increases again as $R$ increases.
The difference in $n_6$ between $R=0$ and $1$ states~(both of which consist of identical size particles), indicates the relatively worse ordering in $R=1$ than $R=0$.
As mentioned above, this type of asymmetry in $n_6$ curves probably comes from the enhanced individual variation and boundary effects. 
As a consequence of this asymmetry, the curve exhibits slight distortion, making it difficult to clearly identify the minimum of $n_6$ exactly at $R=0.5$. By considering the data scattering, we consider the minimum of $n_6$ is achieved at $R \approx 0.5$.

\begin{figure}
\includegraphics[width = 90mm]{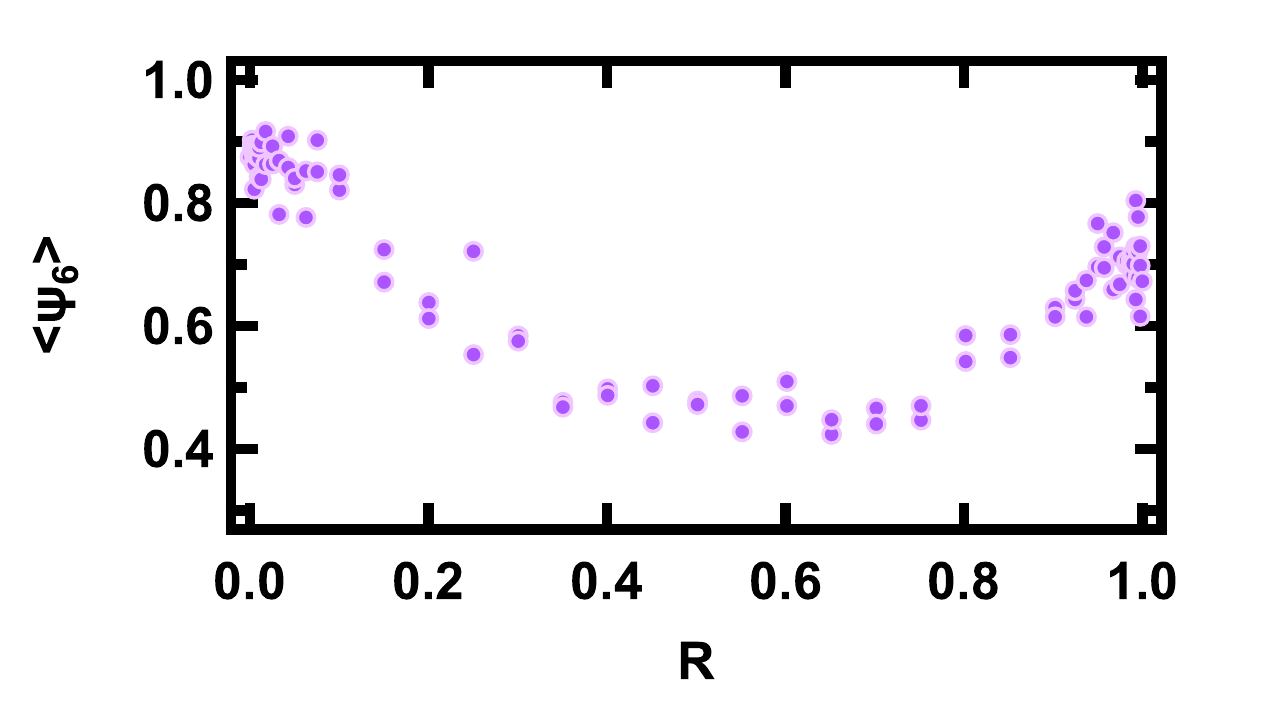}
\caption{$\langle \Psi_6 \rangle$ is plotted as a function of the mixing ratio $R$. Qualitative behavior of $\langle \Psi_6 \rangle$ is very similar to $n_6$ shown in Fig.~\ref{fig:poly_ratio}.}
\label{fig:Psi_6}
\end{figure}

The ratios of pentagonal and heptagonal cells ($n_5$ and $n_7$) are very small at $R=0$ and $1$ (monodisperse states). 
As $R$ deviates from $0$ or $1$, $n_5$ and $n_7$ increase and show the maximum values at $R\approx 0.5$. 
The shapes of $n_5$ and $n_7$ curves are quite similar and they have clear anti-correlation to $n_6$. 
This tendency suggests that the hexagonal ordered structure is disrupted by the defect characterized by the pair of pentagonal and heptagonal cells. 
In this sense, disordering process constructed by bidispersity could be similar to the two-dimensional melting situation. 

To further evaluate this tendency, the $R$ dependence of $\langle \Psi_6 \rangle$ is measured and plotted in FIG.~\ref{fig:Psi_6}. As seen in FIG.~\ref{fig:Psi_6}, the behavior of $\langle \Psi_6(R) \rangle$ is similar to that of $n_6(R)$. This correspondence is natural since both $\langle \Psi_6 \rangle$ and $n_6$ characterize the global hexagonal degree. 
Actually, similar trend of $\langle \Psi_6 \rangle$ behavior has been found in Fig.~7.3 of Ref.\cite{Schockmel:2019}. 

\begin{figure}
\includegraphics[width = 90mm]{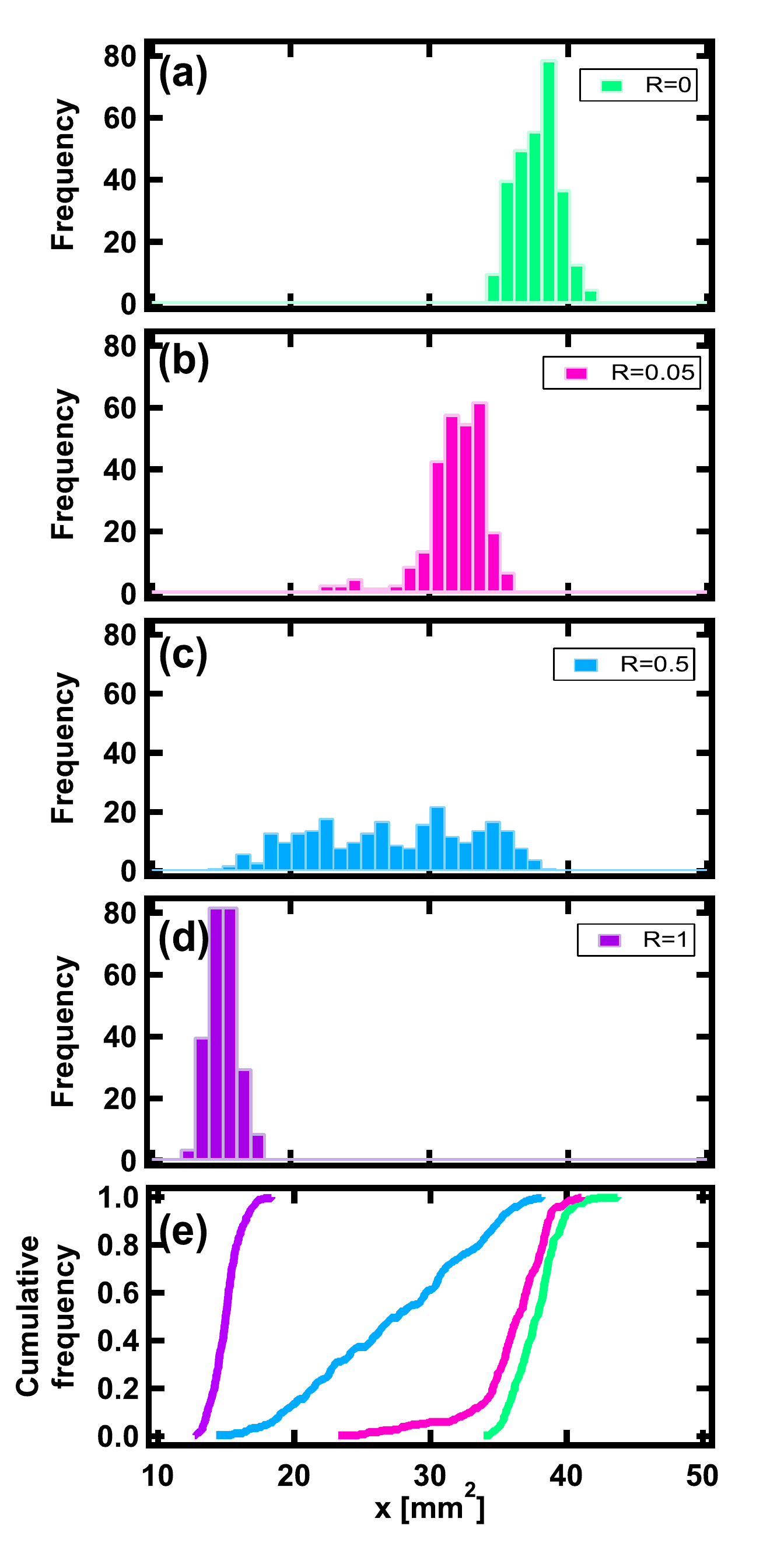}
\caption{\label{fig:Voronoi_area} The distributions of the area of Voronoi cells, where the horizontal axis is the Voronoi cell area and the vertical axis is the frequency. (a) $R=0$, (b) $R=0.05$, (c) $R=0.5$, (d) $R=1$. (e)~Corresponding cumulative distribution curves. Color code used in (e) corresponds to the $R$ values as labeled in (a)-(d) are displayed.}
\end{figure}

For characterizing the degree of ordering in more detail, the dependence of the area distribution of the Voronoi cells on $R$ is examined.
Let $x$ be the Voronoi cell area.
FIG.~\ref{fig:Voronoi_area} shows histograms of the $x$ distribution.
It can be seen that the average of $x$ becomes smaller as $R$ increases.
At $R = 0$ (FIG.~\ref{fig:Voronoi_area}(a)), the $x$ distribution is narrower and symmetric.
At $R$ = 0.05 (FIG.~\ref{fig:Voronoi_area}(b)), the area distribution shifts toward the smaller $x$ range and becomes broader than the distribution at $R = 0$. 
Moreover, the asymmetry of the distribution is enhanced particularly due to the existence of small $x~(\approx \SI{25}{mm^2})$ cells.
This smaller $x$ population represents the small particles contribution.
At $R$ = 0.5 (FIG.~\ref{fig:Voronoi_area}(c)), the distribution of $x$ shifts further toward the smaller $x$ range.
The peak height is very small and the distribution becomes extremely broad.
However, symmetry of the distribution is recovered in this state.
At $R$=1 (FIG.~\ref{fig:Voronoi_area}(d)), the distribution shifts to an even smaller range.
However, the symmetry of the distribution is not disrupted because all the particles in this state are small. 
To clearly show the abovementioned characteristics, the corresponding cumulative distribution curves are also shown in FIG.~\ref{fig:Voronoi_area}(e).

\begin{figure}
\includegraphics[width = 90mm]{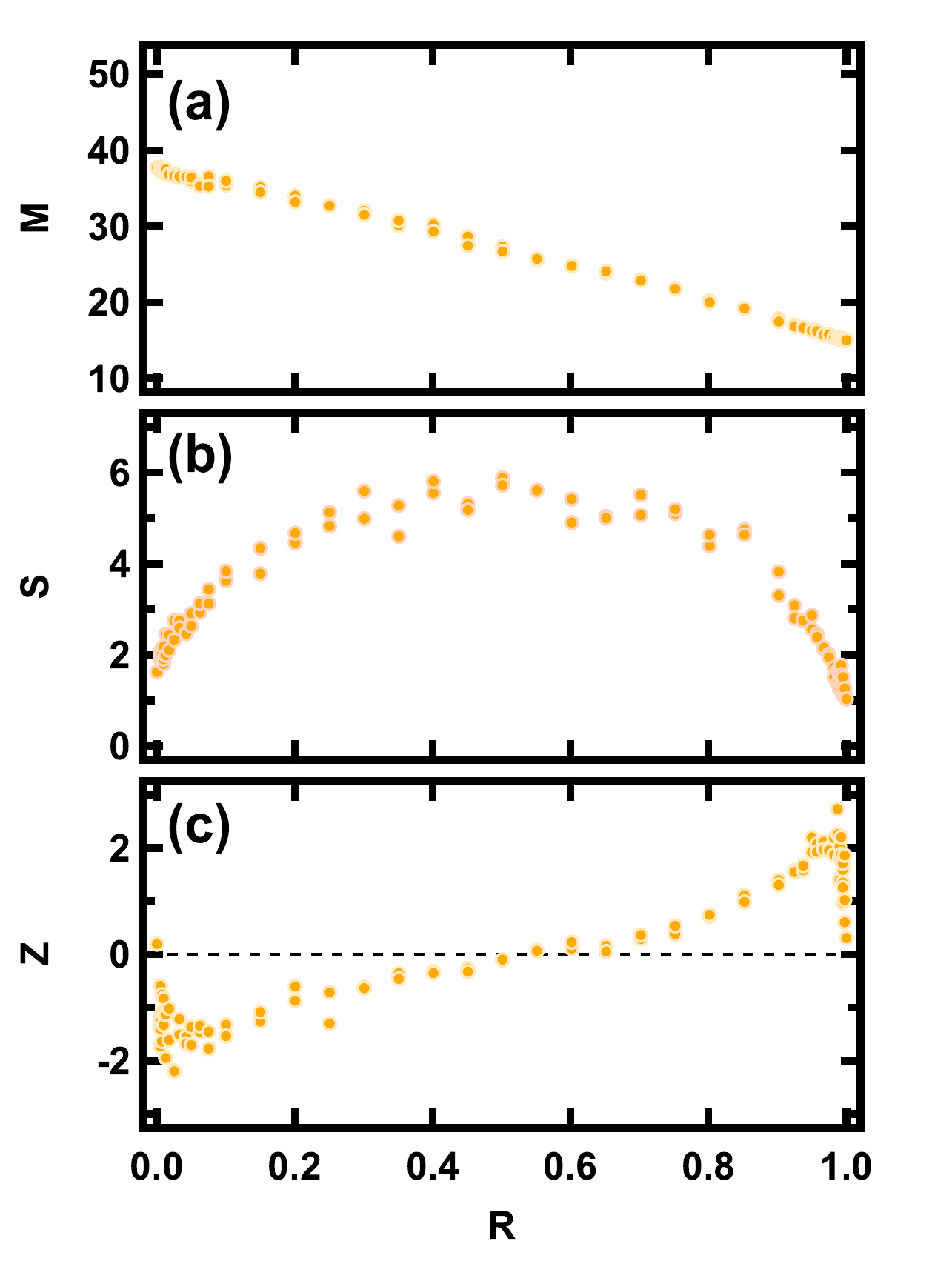}
\caption{\label{fig:moments} Moments of the $x$ (Voronoi cell area) distribution. (a) The mean area of Voronoi cell $M$, (b) the standard deviation $S$, and (c) the skewness $Z$ as functions of the mixture ratio $R$ are presented.}
\end{figure}

In order to characterize the distribution, we evaluate the mean $M$, standard deviation $S$, and skewness $Z$ of the $x$ distribution. 
FIG.~\ref{fig:moments} shows the dependence of $M$, $S$, and $Z$ of the Voronoi cell area on $R$.
Let 
\begin{equation}
m_k=\frac{1}{N}\sum(x-M)^{k}
\end{equation}
be the $k$-th moment of the distribution.
Then, the standard deviation is written as, 
\begin{equation}S=m_2^{1/2}.
\end{equation}
And skewness is written as, 
\begin{equation}
Z=\frac{m_3}{m_2^{3/2}}.
\end{equation}

\begin{figure}
\includegraphics[width = 90mm]{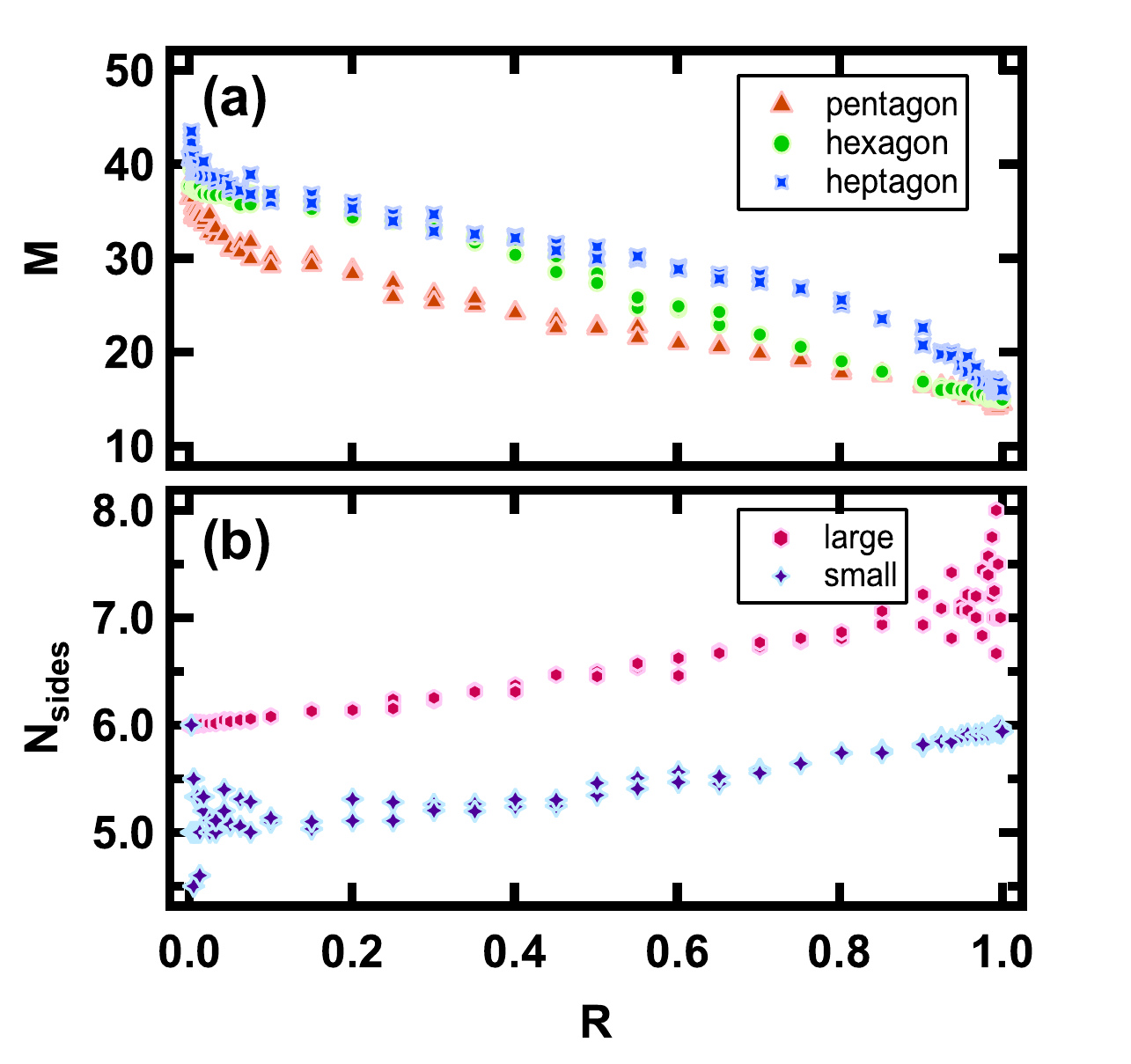}
\caption{\label{fig:l_s}Properties of polygons of the large and small particles. (a) The mean area of Voronoi cell $M$ and (b) the average number of sides of Voronoi cells $N_\mathrm{sides}$ around large and small particles are plotted as a function of mixing ratio $R$.}
\end{figure}

FIG.~\ref{fig:moments}(a) shows that $M$ decreases almost linearly with increasing $R$.
In FIG.~\ref{fig:l_s}(a), average areas of the cells of pentagons, hexagons, and heptagons are shown. As seen in FIG.~\ref{fig:l_s}(a), $M$ of hexagon is close to $M$ of heptagon when $R<0.5$. And when $R>0.5$, $M$ of hexagon approaches $M$ of pentagon. This means that the small particles tend to form smaller pentagons and the large particles tend to form larger heptagons. To clearly show this trend, the relation between the average number of sides (neighbor particles) $N_\mathrm{sides}$ of the Voronoi cells around large or small particles and the mixing ratio $R$ is presented in FIG.~\ref{fig:l_s}. By adding large (or small) particles, a heptagon (or a pentagon) is formed around the added particle. By this disturbance, some major small (or large) particle's Voronoi cell becomes a pentagon (or a heptagon). The two-dimensional hexagonal order is disrupted by introducing bidispersity in this manner. 

In FIG.~\ref{fig:moments}(b), $S$ has a small value at $R=0$ and increases as $R$ increases.
Then, $S$ shows the maximum value around $R=0.5$.
In the range of $R > 0.5$, $S$ decreases as $R$ increases.
The shape of $S(R)$ is qualitatively similar to $n_{5,7}$ curves and almost symmetric. 
If $S$ is normalized to $M$, $S/M$ curve shows a slightly asymmetric feature like $n_{5,7}$ curves. 

In FIG.~\ref{fig:moments}(c), one can confirm that $Z$ is approximately $0$ at $R=0$.
Then, the value of $Z$ rapidly decreases with increasing $R$ until $R\approx 0.05$ where $Z$ has a negative peak~(minimum) value.
Negative values of $Z$ indicate that the distribution is asymmetric due to the tail in the small $x$ range.
This asymmetry of the distribution characterized by $Z$ is enhanced when approximately 5\% of small particles are present in the system.
Then, in the range of $0.05 \lesssim R \lesssim 0.95$, $Z$ increases as $R$ increases.
At $R=0.5$, $Z$ becomes almost 0.
This indicates that $x$ is symmetrically distributed at $R=0.5$.
$Z$ shows the maximum value around $R=0.95$.
This implies that the presence of a small number ($\approx 5\%$) of a large Voronoi cells yields the asymmetry in the $x$ distribution, as in the case of $R=0.05$.

\section{Discussion}

First, we compare our results with the previous study\cite{forcechain} that used the photoelastic particles to investigate the relation between mixing ratio and the orientational order of force chain structure in two-dimensional granular system.
In this previous study, the degree of orientational order shows the maximum when $R=0$.
Then, as $R$ increases, the value of orientational order decreases until around $R=0.1$.
However, the value of orientational order does not vary in the range of $0.1 \leq R \leq0.9$.
From $R=0.9$ to $R=1$, the value of orientational order increases with increasing $R$. 
At $R=1$, the value of orientational order again shows the maximum value.
From these results,  the previous study concluded that about 10\% bidispersity is sufficient to achieve the spatially homogeneous random structure.
However, the current experimental results present a different scenario to develop random structure by bidispersity.
The characteristic feature found in this study is the continuous decrease of $n_6(R)$ and $\langle \Psi_6(R) \rangle$ from $R=0$ to $R=0.5$. 
We do not observe clear plateau behavior in the range of $0.1 \leq R \leq 0.9$.
In addition, in FIG.~\ref{fig:moments}(a), $M(R)$ linearly decreases from $R=0$ to $R=1$.
In FIG.~\ref{fig:moments}(b), $S$ continuously increases from $R=0$ to $R=0.5$. 
Regarding the skewness behavior, $Z(R)$ shows a peak value at $R\approx 0.05$ as shown in FIG.~\ref{fig:moments}(d).
However, $Z(R)$ continuously varies in $R\geq0.05.$
The symmetric $x$ distribution characterized by $Z\approx 0$ is achieved at $R\approx 0.5$ and the $Z$ value varies symmetrically around $R\approx 0.5$. 
Thus, in the present magnet-particle system, we consider that the degree of configurational order continuously vary all the way from $R=0$ to $R=0.5$. Perhaps, the best mixing ratio ($R\approx 0.5$ in this study) might depend on the size ratio of large and small particles. Although we guess $R\approx 0.5$ is appropriate for realistic size-ratio range (e.g. less than factor 2 difference), size ratio dependence of the best $R$ value is a future problem. 

The current result suggests that $R\approx 0.5$ is an ideal mixing of large and small particles to achieve the symmetric random distribution. As revealed by the previous study~\cite{forcechain}, $10\%$ bidispersity is sufficient to make random structures in terms of force chain orientational order. 
Our study reveals that the skewness suggests that the ordered structure is more or less disrupted when $R\approx 5$\%. This significant skewness can be attributed to the contribution of small particles to the disordering of the particle configuration. By considering Voronoi-cell-area distribution and global bond orientational order, we find the maximum disorder is attained at $R\approx 0.5$. However, note that we do not measure the force chain network. 
It is difficult to define force chain structure in the magnetic particle system. Thus, we characterize the particle configuration directly.
In the current study, asymmetry of the Voronoi-cell-area distribution can surely be confirmed at $10\%$ bidisperse state.
To eliminate the asymmetry, $\approx 50\%$ bidispersity is better. If one aims to mimic natural disorder systems in granular experiments/simulations using bidispersity, using only 10\% bidispersity may not suffice. This is due to the asymmetric distribution characterized by the skewness, which can affect the macroscopic behaviors of the system. To eliminate such artificial effects, using bidispersity of approximately 50\% is better. 
Furthermore, the ratios among pentagonal, hexagonal, and heptagonal structures also indicate that $\approx $50\% bidispersity $(R\approx 0.5)$ is better to form a random structure. However, both the interparticle contact force and analysis method are different between the previous work~\cite{forcechain} and current study. 
The effective force chain structure might be defined in magnetic system as well. Quantitative analysis of the force balance in the magnet-particles system might reveal the details of the particle configuration. 
Further systematic experiments are necessary to clarify details about the effect of physical contact among particles. 
This is an important future work.

Usually, the spatial correlation of bond orientational order $G_6(|\bm{r}-\bm{r'}|) = \langle \exp \left(6i \left[ \theta(\bm{r}) - \theta(\bm{r'}) \right] \right) \rangle$, where $\bm{r}$ and $\bm{r'}$ are the position vectors of the particles and $\theta$ denotes the angle to the horizontal axis, is also used to characterize the ordered degree, this value does not show clear $R$ dependent variation in the current system. $G_6(r)$ data scatter and it is difficult to clearly confirm the $r \, (=|\bm{r}-\bm{r'}|)$-dependent decaying behavior. Only the global bond order parameter $\langle\Psi_6 \rangle$ can properly characterize the system in this study. 
This difficulty comes from the less number of particles and very compressed (thin) cell's anisotropic effects including boundary walls. Moreover, we realize that the systematic analysis of radial distribution function is also difficult due to the same reason. 
Namely, although the number of particles in the current experimental system is small and the system is thin to directly evaluate through spatial correlations such as $G_6(r)$ and radial distribution function, the statistics of Voronoi cells and $\langle \Psi_6 \rangle$ allow us to characterize the global degree of the ordering in two-dimensional bidisperse system. The shape and area distributions of the Voronoi cells can characterize the state of randomness. 

\begin{figure}
\includegraphics[width = 90mm]{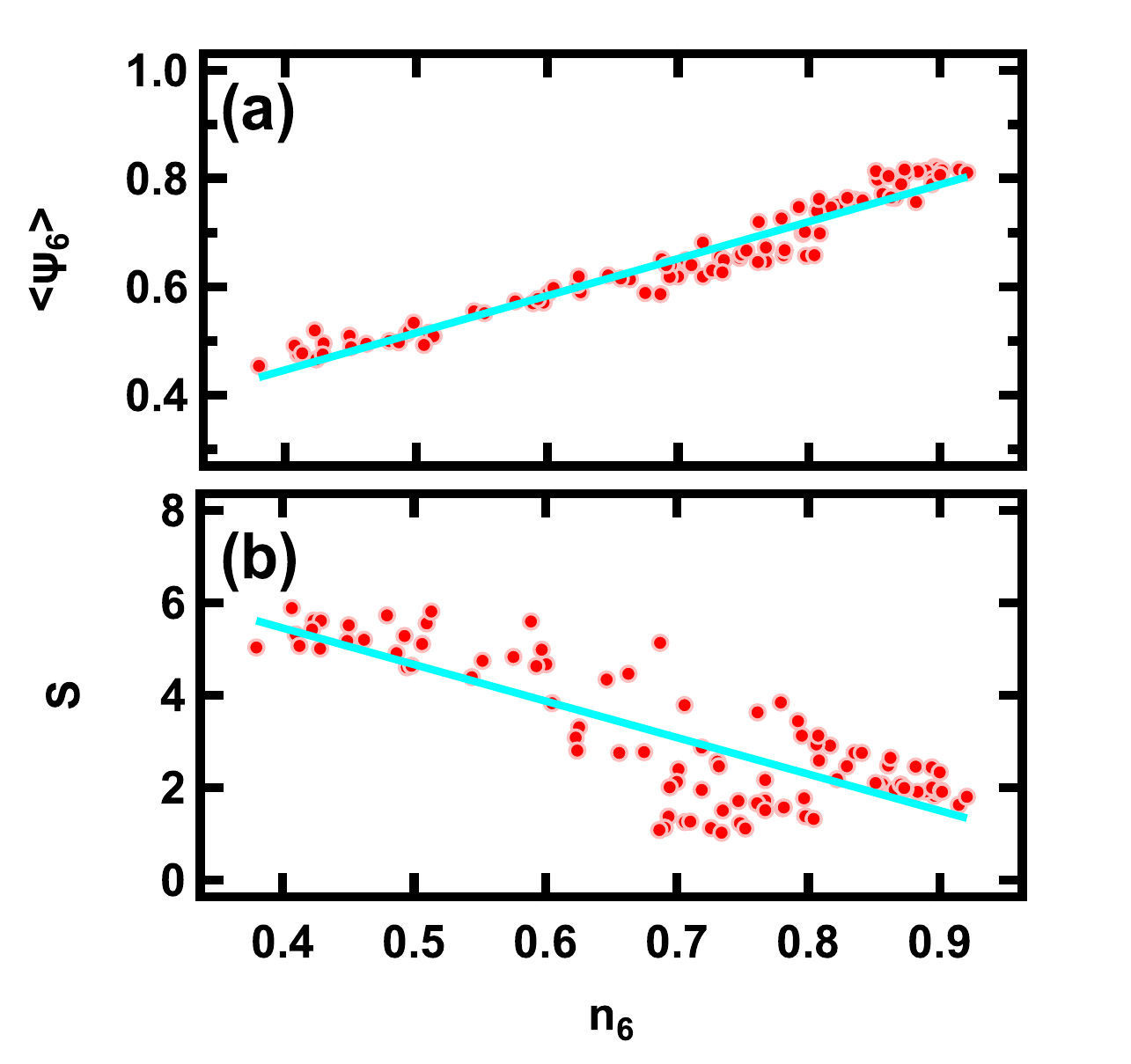}
\caption{\label{fig:corr} Correlations between (a)~$\langle \Psi_6 \rangle$ vs.~$n_6$ and (b)~$S$ vs.~$n_6$. In both panels, clear correlations can be confirmed as shown by the linear fitting lines. Correlation coefficients are $0.97$ and $-0.81$ for (a) and (b), respectively.}
\end{figure}

Based on Figs.~\ref{fig:poly_ratio}, \ref{fig:Psi_6}, and \ref{fig:moments}(b), it can be observed that $n_6$, $\langle \Psi_6 \rangle$, and $S$ exhibit qualitatively similar trends. To quantitatively characterize this relationship, direct correlations among $n_6$, $\langle \Psi_6 \rangle$, and $S$ are shown in Fig.~\ref{fig:corr}. One can confirm the good correlations. Correlation coefficients are $0.97$ and $-0.81$ for $\langle \Psi_6 \rangle$ vs.~$n_6$ and $S$ vs.~$n_6$, respectively. Thus, we consider any of them can similarly characterize the disorder degree. 

In the current system, we do not have any thermal effect such as random motion of particles. 
To mimic the two-dimensional melting by this system, a certain analogue of thermal fluctuation should be added to the system. 
Perhaps, mechanical vibration can be used as a source of such fluctuation~\cite{Schockmel:2013}. 
Then, we can consider the melting of the ordered structure by the thermal effect. 
The temperature is directly measurable by the random motion of the particles in this system. 
This is the advantage of a macroscopic granular system. 
The pressure is certainly measurable by the confining piston force. 
Then, the equation of state of the magnetic particle system can be built through this kind of experiment. 
Finally, slow compression force of the monodisperse magnet particle system was reported by Modesto et al.~\cite{compression}
The effect of bidispersity on the compression force would also be an interesting future problem.

Moreover, our analysis is limited to the highly compressed state achieved with 110 N compression force. It has been reported in Ref.\cite{compression} that the compression force nonlinearly increases as the compression proceeds and exhibits hysteresis. The relation between particle configuration order and compression force is an intriguing topic for further exploration. By increasing the compression force, the effect of gravity on the particle configuration order would be gradually reduced. Investigating the structural development of the compressed magnet system in more detail is also an interesting future research problem. Additionally, the effect of bidispersity on compression force can be examined by comparing our experimental results with the previous study. To eliminate the effect of gravity, a horizontal setup can be employed, which is currently being pursued as an extension of our experimental setup. It should be noted that the gravity effect is negligible even in the current vertical setup as long as the granular layer is sufficiently compressed. If the horizontal setup is used, the gravity effect becomes negligible from the initial state prior to the compression. In addition, we plan to perform corresponding numerical simulations to investigate a wider range of parameter values. The results of these investigations will be reported in future publications.

\section{Conclusion}

The  effect of bidispersity on the ordered particle configuration is experimentally studied in a two-dimensional granular layer consisting of repulsive magnets. 
Voronoi tessellation is used to investigate the $R$ (mixing ratio) dependence of the degree of ordering through the hexagonal ratio of Voronoi cells $n_6$, the global bond orientational order parameter $\langle \Psi_6 \rangle$, and the Voronoi cell area distribution.
As a result, $n_6$ and $S$ (standard deviation of area distribution) show the continuous peak around $R\approx 0.5$.
The value of $Z$ (skewness of area distribution) shows the peak at $R\approx 0.05$, and $Z$ becomes almost $0$ at $R\approx 0.5$. These results indicate that the random structure without any asymmetric distribution is achieved at $R\approx 0.5$. This is in contrast with the results of previous study, which showed that the value of orientational order of force chain structure does not change in the range of $R=0.1$--$0.9$. According to the current result, a truly equal number bidispersity is favorable to attain the ideal random structure by using bidispersity.
The system we use in this study is different from a usual granular system.
Therefore, further studies with various setups are necessary to discuss general features to achieve an ideally random granular structure.

\section*{Acknowledgement}
This work was partially supported by JSPS KAKENHI Grant, No.~18H03679. 
S.D. thanks F.R.S.-FNRS for financial support as senior research associate. 
Y.D.S. thanks FAP-DR Project 00193-00001155/2021-40 for financial support.

\section*{Author Declarations}
\subsection*{Conflict of Interest}
The authors have no conflicts to disclose.
\subsection*{Author Contributions}
\textbf{K. Tsuchikusa}: Conceptualization (equal); Investigation (lead); Data curation (lead); Writing – original draft (lead); Writing – review \& editing (supporting). 
\textbf{K. Yamamoto}: Supervision (supporting); Investigation (supporting); Writing – review \& editing (supporting). 
\textbf{M. Katsura}: Supervision (supporting); Investigation (supporting); Writing – review \& editing (supporting). 
\textbf{C.T. de Paula}: Conceptualization (equal); Writing – review \& editing (supporting). 
\textbf{J.A.C. Modesto}: Conceptualization (equal); Writing – review \& editing (supporting). 
\textbf{S. Dorbolo}: Conceptualization (equal); Writing – review \& editing (supporting). 
\textbf{F. Pacheco-V\'azquez}: Conceptualization (equal); Writing – review \& editing (supporting). 
\textbf{Y.D. Sobral}: Conceptualization (equal); Writing – review \& editing (supporting). 
\textbf{H. Katsuragi}: Conceptualization (equal); Supervision (lead); Investigation (supporting);  Writing – original draft (supporting); Writing – review \& editing (lead). 

\section*{Data Availability}
The data that support the findings of this study are available from the corresponding author upon reasonable request.

\nocite{*}
\bibliography{tsuchi}

\providecommand{\noopsort}[1]{}\providecommand{\singleletter}[1]{#1}%
\begin{thebibliography}{31}%
\makeatletter
\providecommand \@ifxundefined [1]{%
 \@ifx{#1\undefined}
}%
\providecommand \@ifnum [1]{%
 \ifnum #1\expandafter \@firstoftwo
 \else \expandafter \@secondoftwo
 \fi
}%
\providecommand \@ifx [1]{%
 \ifx #1\expandafter \@firstoftwo
 \else \expandafter \@secondoftwo
 \fi
}%
\providecommand \natexlab [1]{#1}%
\providecommand \enquote  [1]{``#1''}%
\providecommand \bibnamefont  [1]{#1}%
\providecommand \bibfnamefont [1]{#1}%
\providecommand \citenamefont [1]{#1}%
\providecommand \href@noop [0]{\@secondoftwo}%
\providecommand \href [0]{\begingroup \@sanitize@url \@href}%
\providecommand \@href[1]{\@@startlink{#1}\@@href}%
\providecommand \@@href[1]{\endgroup#1\@@endlink}%
\providecommand \@sanitize@url [0]{\catcode `\\12\catcode `\$12\catcode
  `\&12\catcode `\#12\catcode `\^12\catcode `\_12\catcode `\%12\relax}%
\providecommand \@@startlink[1]{}%
\providecommand \@@endlink[0]{}%
\providecommand \url  [0]{\begingroup\@sanitize@url \@url }%
\providecommand \@url [1]{\endgroup\@href {#1}{\urlprefix }}%
\providecommand \urlprefix  [0]{URL }%
\providecommand \Eprint [0]{\href }%
\providecommand \doibase [0]{http://dx.doi.org/}%
\providecommand \selectlanguage [0]{\@gobble}%
\providecommand \bibinfo  [0]{\@secondoftwo}%
\providecommand \bibfield  [0]{\@secondoftwo}%
\providecommand \translation [1]{[#1]}%
\providecommand \BibitemOpen [0]{}%
\providecommand \bibitemStop [0]{}%
\providecommand \bibitemNoStop [0]{.\EOS\space}%
\providecommand \EOS [0]{\spacefactor3000\relax}%
\providecommand \BibitemShut  [1]{\csname bibitem#1\endcsname}%
\let\auto@bib@innerbib\@empty
\bibitem [{\citenamefont {Iikawa}, \citenamefont {Bandi},\ and\ \citenamefont
  {Katsuragi}(2016)}]{forcechain}%
  \BibitemOpen
  \bibfield  {author} {\bibinfo {author} {\bibfnamefont {N.}~\bibnamefont
  {Iikawa}}, \bibinfo {author} {\bibfnamefont {M.~M.}\ \bibnamefont {Bandi}}, \
  and\ \bibinfo {author} {\bibfnamefont {H.}~\bibnamefont {Katsuragi}},\
  }\bibfield  {title} {\enquote {\bibinfo {title} {{Sensitivity of Granular
  Force Chain Orientation to Disorder-Induced Metastable Relaxation}},}\ }\href
  {\doibase 10.1103/physrevlett.116.128001} {\bibfield  {journal} {\bibinfo
  {journal} {Physical Review Letters}\ }\textbf {\bibinfo {volume} {116}},\
  \bibinfo {pages} {128001} (\bibinfo {year} {2016})}\BibitemShut {NoStop}%
\bibitem [{\citenamefont {Reinhardt}\ \emph {et~al.}(2017)\citenamefont
  {Reinhardt}, \citenamefont {Schreck}, \citenamefont {Romano},\ and\
  \citenamefont {Doye}}]{Reinhardt:2017}%
  \BibitemOpen
  \bibfield  {author} {\bibinfo {author} {\bibfnamefont {A.}~\bibnamefont
  {Reinhardt}}, \bibinfo {author} {\bibfnamefont {J.~S.}\ \bibnamefont
  {Schreck}}, \bibinfo {author} {\bibfnamefont {F.}~\bibnamefont {Romano}}, \
  and\ \bibinfo {author} {\bibfnamefont {J.~P.~K.}\ \bibnamefont {Doye}},\
  }\bibfield  {title} {\enquote {\bibinfo {title} {{Self-assembly of
  two-dimensional binary quasicrystals: a possible route to a DNA
  quasicrystal}},}\ }\href {\doibase 10.1088/0953-8984/29/1/014006} {\bibfield
  {journal} {\bibinfo  {journal} {Journal of Physics: Condensed Matter}\
  }\textbf {\bibinfo {volume} {29}},\ \bibinfo {pages} {014006} (\bibinfo
  {year} {2017})},\ \Eprint {http://arxiv.org/abs/1607.06626} {1607.06626}
  \BibitemShut {NoStop}%
\bibitem [{\citenamefont {Fayen}\ \emph {et~al.}(2020)\citenamefont {Fayen},
  \citenamefont {Jagannathan}, \citenamefont {Foffi},\ and\ \citenamefont
  {Smallenburg}}]{Fayen:2020}%
  \BibitemOpen
  \bibfield  {author} {\bibinfo {author} {\bibfnamefont {E.}~\bibnamefont
  {Fayen}}, \bibinfo {author} {\bibfnamefont {A.}~\bibnamefont {Jagannathan}},
  \bibinfo {author} {\bibfnamefont {G.}~\bibnamefont {Foffi}}, \ and\ \bibinfo
  {author} {\bibfnamefont {F.}~\bibnamefont {Smallenburg}},\ }\bibfield
  {title} {\enquote {\bibinfo {title} {{Infinite-pressure phase diagram of
  binary mixtures of (non)additive hard disks}},}\ }\href {\doibase
  10.1063/5.0008230} {\bibfield  {journal} {\bibinfo  {journal} {The Journal of
  Chemical Physics}\ }\textbf {\bibinfo {volume} {152}},\ \bibinfo {pages}
  {204901} (\bibinfo {year} {2020})},\ \Eprint
  {http://arxiv.org/abs/2003.08889} {2003.08889} \BibitemShut {NoStop}%
\bibitem [{\citenamefont {Fayen}\ \emph {et~al.}(2022)\citenamefont {Fayen},
  \citenamefont {Impéror-Clerc}, \citenamefont {Filion}, \citenamefont
  {Foffi},\ and\ \citenamefont {Smallenburg}}]{Fayen:2022}%
  \BibitemOpen
  \bibfield  {author} {\bibinfo {author} {\bibfnamefont {E.}~\bibnamefont
  {Fayen}}, \bibinfo {author} {\bibfnamefont {M.}~\bibnamefont
  {Impéror-Clerc}}, \bibinfo {author} {\bibfnamefont {L.}~\bibnamefont
  {Filion}}, \bibinfo {author} {\bibfnamefont {G.}~\bibnamefont {Foffi}}, \
  and\ \bibinfo {author} {\bibfnamefont {F.}~\bibnamefont {Smallenburg}},\
  }\bibfield  {title} {\enquote {\bibinfo {title} {{Self-assembly of
  dodecagonal and octagonal quasicrystals in hard spheres on a plane}},}\
  }\href@noop {} {\bibfield  {journal} {\bibinfo  {journal} {arXiv}\ }
  (\bibinfo {year} {2022})},\ \Eprint {http://arxiv.org/abs/2202.12726}
  {2202.12726} \BibitemShut {NoStop}%
\bibitem [{\citenamefont {Assoud}, \citenamefont {Messina},\ and\ \citenamefont
  {L\"owen}(2007)}]{Assoud:2007}%
  \BibitemOpen
  \bibfield  {author} {\bibinfo {author} {\bibfnamefont {L.}~\bibnamefont
  {Assoud}}, \bibinfo {author} {\bibfnamefont {R.}~\bibnamefont {Messina}}, \
  and\ \bibinfo {author} {\bibfnamefont {H.}~\bibnamefont {L\"owen}},\
  }\bibfield  {title} {\enquote {\bibinfo {title} {{Stable crystalline lattices
  in two-dimensional binary mixtures of dipolar particles}},}\ }\href {\doibase
  10.1209/0295-5075/80/48001} {\bibfield  {journal} {\bibinfo  {journal} {EPL
  (Europhysics Letters)}\ }\textbf {\bibinfo {volume} {80}},\ \bibinfo {pages}
  {48001} (\bibinfo {year} {2007})},\ \Eprint {http://arxiv.org/abs/0706.2311}
  {0706.2311} \BibitemShut {NoStop}%
\bibitem [{\citenamefont {Messina}\ and\ \citenamefont
  {Aljawhari}(2016)}]{Messina:2016}%
  \BibitemOpen
  \bibfield  {author} {\bibinfo {author} {\bibfnamefont {R.}~\bibnamefont
  {Messina}}\ and\ \bibinfo {author} {\bibfnamefont {S.}~\bibnamefont
  {Aljawhari}},\ }\bibfield  {title} {\enquote {\bibinfo {title}
  {{Crystallization of binary mixtures of similar dipole moments in two
  dimensions: A Monte Carlo study}},}\ }\href {\doibase
  10.1209/0295-5075/115/28005} {\bibfield  {journal} {\bibinfo  {journal}
  {Europhysics Letters}\ }\textbf {\bibinfo {volume} {115}},\ \bibinfo {pages}
  {28005} (\bibinfo {year} {2016})}\BibitemShut {NoStop}%
\bibitem [{\citenamefont {Ebert}, \citenamefont {Keim},\ and\ \citenamefont
  {Maret}(2008)}]{Ebert:2008}%
  \BibitemOpen
  \bibfield  {author} {\bibinfo {author} {\bibfnamefont {F.}~\bibnamefont
  {Ebert}}, \bibinfo {author} {\bibfnamefont {P.}~\bibnamefont {Keim}}, \ and\
  \bibinfo {author} {\bibfnamefont {G.}~\bibnamefont {Maret}},\ }\bibfield
  {title} {\enquote {\bibinfo {title} {{Local crystalline order in a 2D
  colloidal glass former}},}\ }\href {\doibase 10.1140/epje/i2007-10270-8}
  {\bibfield  {journal} {\bibinfo  {journal} {The European Physical Journal E}\
  }\textbf {\bibinfo {volume} {26}},\ \bibinfo {pages} {161--168} (\bibinfo
  {year} {2008})}\BibitemShut {NoStop}%
\bibitem [{\citenamefont {Fornleitner}\ \emph {et~al.}(2008)\citenamefont
  {Fornleitner}, \citenamefont {Verso}, \citenamefont {Kahl},\ and\
  \citenamefont {Likos}}]{Fornleitner:2008}%
  \BibitemOpen
  \bibfield  {author} {\bibinfo {author} {\bibfnamefont {J.}~\bibnamefont
  {Fornleitner}}, \bibinfo {author} {\bibfnamefont {F.~L.}\ \bibnamefont
  {Verso}}, \bibinfo {author} {\bibfnamefont {G.}~\bibnamefont {Kahl}}, \ and\
  \bibinfo {author} {\bibfnamefont {C.~N.}\ \bibnamefont {Likos}},\ }\bibfield
  {title} {\enquote {\bibinfo {title} {{Genetic algorithms predict formation of
  exotic ordered configurations for two-component dipolar monolayers}},}\
  }\href {\doibase 10.1039/b717205b} {\bibfield  {journal} {\bibinfo  {journal}
  {Soft Matter}\ }\textbf {\bibinfo {volume} {4}},\ \bibinfo {pages} {480--484}
  (\bibinfo {year} {2008})}\BibitemShut {NoStop}%
\bibitem [{\citenamefont {Fornleitner}\ \emph {et~al.}(2009)\citenamefont
  {Fornleitner}, \citenamefont {Verso}, \citenamefont {Kahl},\ and\
  \citenamefont {Likos}}]{Fornleitner:2009}%
  \BibitemOpen
  \bibfield  {author} {\bibinfo {author} {\bibfnamefont {J.}~\bibnamefont
  {Fornleitner}}, \bibinfo {author} {\bibfnamefont {F.~L.}\ \bibnamefont
  {Verso}}, \bibinfo {author} {\bibfnamefont {G.}~\bibnamefont {Kahl}}, \ and\
  \bibinfo {author} {\bibfnamefont {C.~N.}\ \bibnamefont {Likos}},\ }\bibfield
  {title} {\enquote {\bibinfo {title} {{Ordering in Two-Dimensional Dipolar
  Mixtures}},}\ }\href {\doibase 10.1021/la900421v} {\bibfield  {journal}
  {\bibinfo  {journal} {Langmuir}\ }\textbf {\bibinfo {volume} {25}},\ \bibinfo
  {pages} {7836--7846} (\bibinfo {year} {2009})}\BibitemShut {NoStop}%
\bibitem [{\citenamefont {Schockmel}(2019)}]{Schockmel:2019}%
  \BibitemOpen
  \bibfield  {author} {\bibinfo {author} {\bibfnamefont {J.}~\bibnamefont
  {Schockmel}},\ }\emph {\bibinfo {title} {Self-organization of a monolayer of
  magnetized beads}},\ \href@noop {} {Ph.D. thesis},\ \bibinfo  {school}
  {Universit\'e de Li\`ege} (\bibinfo {year} {2019})\BibitemShut {NoStop}%
\bibitem [{\citenamefont {Ebert}, \citenamefont {Maret},\ and\ \citenamefont
  {Keim}(2009)}]{Ebert:2009}%
  \BibitemOpen
  \bibfield  {author} {\bibinfo {author} {\bibfnamefont {F.}~\bibnamefont
  {Ebert}}, \bibinfo {author} {\bibfnamefont {G.}~\bibnamefont {Maret}}, \ and\
  \bibinfo {author} {\bibfnamefont {P.}~\bibnamefont {Keim}},\ }\bibfield
  {title} {\enquote {\bibinfo {title} {{Partial clustering prevents global
  crystallization in a binary 2D colloidal glass former}},}\ }\href {\doibase
  10.1140/epje/i2009-10490-x} {\bibfield  {journal} {\bibinfo  {journal} {The
  European Physical Journal E}\ }\textbf {\bibinfo {volume} {29}},\ \bibinfo
  {pages} {311--318} (\bibinfo {year} {2009})},\ \Eprint
  {http://arxiv.org/abs/0903.2812} {0903.2812} \BibitemShut {NoStop}%
\bibitem [{\citenamefont {Assoud}\ \emph {et~al.}(2009)\citenamefont {Assoud},
  \citenamefont {Ebert}, \citenamefont {Keim}, \citenamefont {Messina},
  \citenamefont {Maret},\ and\ \citenamefont {L\"owen}}]{Assoud:2009}%
  \BibitemOpen
  \bibfield  {author} {\bibinfo {author} {\bibfnamefont {L.}~\bibnamefont
  {Assoud}}, \bibinfo {author} {\bibfnamefont {F.}~\bibnamefont {Ebert}},
  \bibinfo {author} {\bibfnamefont {P.}~\bibnamefont {Keim}}, \bibinfo {author}
  {\bibfnamefont {R.}~\bibnamefont {Messina}}, \bibinfo {author} {\bibfnamefont
  {G.}~\bibnamefont {Maret}}, \ and\ \bibinfo {author} {\bibfnamefont
  {H.}~\bibnamefont {L\"owen}},\ }\bibfield  {title} {\enquote {\bibinfo
  {title} {{Ultrafast Quenching of Binary Colloidal Suspensions in an External
  Magnetic Field}},}\ }\href {\doibase 10.1103/physrevlett.102.238301}
  {\bibfield  {journal} {\bibinfo  {journal} {Physical Review Letters}\
  }\textbf {\bibinfo {volume} {102}},\ \bibinfo {pages} {238301} (\bibinfo
  {year} {2009})},\ \Eprint {http://arxiv.org/abs/0811.1498} {0811.1498}
  \BibitemShut {NoStop}%
\bibitem [{\citenamefont {Strandburg}(1988)}]{Strandburg:1988}%
  \BibitemOpen
  \bibfield  {author} {\bibinfo {author} {\bibfnamefont {K.~J.}\ \bibnamefont
  {Strandburg}},\ }\bibfield  {title} {\enquote {\bibinfo {title}
  {{Two-dimensional melting}},}\ }\href {\doibase 10.1103/revmodphys.60.161}
  {\bibfield  {journal} {\bibinfo  {journal} {Reviews of Modern Physics}\
  }\textbf {\bibinfo {volume} {60}},\ \bibinfo {pages} {161--207} (\bibinfo
  {year} {1988})}\BibitemShut {NoStop}%
\bibitem [{\citenamefont {Gasser}\ \emph {et~al.}(2010)\citenamefont {Gasser},
  \citenamefont {Eisenmann}, \citenamefont {Maret},\ and\ \citenamefont
  {Keim}}]{Gasser:2010}%
  \BibitemOpen
  \bibfield  {author} {\bibinfo {author} {\bibfnamefont {U.}~\bibnamefont
  {Gasser}}, \bibinfo {author} {\bibfnamefont {C.}~\bibnamefont {Eisenmann}},
  \bibinfo {author} {\bibfnamefont {G.}~\bibnamefont {Maret}}, \ and\ \bibinfo
  {author} {\bibfnamefont {P.}~\bibnamefont {Keim}},\ }\bibfield  {title}
  {\enquote {\bibinfo {title} {{Melting of Crystals in Two Dimensions}},}\
  }\href {\doibase 10.1002/cphc.200900755} {\bibfield  {journal} {\bibinfo
  {journal} {ChemPhysChem}\ }\textbf {\bibinfo {volume} {11}},\ \bibinfo
  {pages} {963--970} (\bibinfo {year} {2010})}\BibitemShut {NoStop}%
\bibitem [{\citenamefont {Schockmel}\ \emph {et~al.}(2013)\citenamefont
  {Schockmel}, \citenamefont {Mersch}, \citenamefont {Vandewalle},\ and\
  \citenamefont {Lumay}}]{Schockmel:2013}%
  \BibitemOpen
  \bibfield  {author} {\bibinfo {author} {\bibfnamefont {J.}~\bibnamefont
  {Schockmel}}, \bibinfo {author} {\bibfnamefont {E.}~\bibnamefont {Mersch}},
  \bibinfo {author} {\bibfnamefont {N.}~\bibnamefont {Vandewalle}}, \ and\
  \bibinfo {author} {\bibfnamefont {G.}~\bibnamefont {Lumay}},\ }\bibfield
  {title} {\enquote {\bibinfo {title} {{Melting of a confined monolayer of
  magnetized beads}},}\ }\href {\doibase 10.1103/physreve.87.062201} {\bibfield
   {journal} {\bibinfo  {journal} {Physical Review E}\ }\textbf {\bibinfo
  {volume} {87}},\ \bibinfo {pages} {062201} (\bibinfo {year}
  {2013})}\BibitemShut {NoStop}%
\bibitem [{\citenamefont {Messina}\ \emph {et~al.}(2015)\citenamefont
  {Messina}, \citenamefont {Aljawhari}, \citenamefont {B\'ecu}, \citenamefont
  {Schockmel}, \citenamefont {Lumay},\ and\ \citenamefont
  {Vandewalle}}]{Messina:2015}%
  \BibitemOpen
  \bibfield  {author} {\bibinfo {author} {\bibfnamefont {R.}~\bibnamefont
  {Messina}}, \bibinfo {author} {\bibfnamefont {S.}~\bibnamefont {Aljawhari}},
  \bibinfo {author} {\bibfnamefont {L.}~\bibnamefont {B\'ecu}}, \bibinfo
  {author} {\bibfnamefont {J.}~\bibnamefont {Schockmel}}, \bibinfo {author}
  {\bibfnamefont {G.}~\bibnamefont {Lumay}}, \ and\ \bibinfo {author}
  {\bibfnamefont {N.}~\bibnamefont {Vandewalle}},\ }\bibfield  {title}
  {\enquote {\bibinfo {title} {{Quantitatively mimicking wet colloidal
  suspensions with dry granular media}},}\ }\href {\doibase 10.1038/srep10348}
  {\bibfield  {journal} {\bibinfo  {journal} {Scientific Reports}\ }\textbf
  {\bibinfo {volume} {5}},\ \bibinfo {pages} {10348} (\bibinfo {year}
  {2015})}\BibitemShut {NoStop}%
\bibitem [{\citenamefont {Schockmel}\ \emph {et~al.}(2017)\citenamefont
  {Schockmel}, \citenamefont {Vandewalle}, \citenamefont {Opsomer},\ and\
  \citenamefont {Lumay}}]{Schockmel:2017}%
  \BibitemOpen
  \bibfield  {author} {\bibinfo {author} {\bibfnamefont {J.}~\bibnamefont
  {Schockmel}}, \bibinfo {author} {\bibfnamefont {N.}~\bibnamefont
  {Vandewalle}}, \bibinfo {author} {\bibfnamefont {E.}~\bibnamefont {Opsomer}},
  \ and\ \bibinfo {author} {\bibfnamefont {G.}~\bibnamefont {Lumay}},\
  }\bibfield  {title} {\enquote {\bibinfo {title} {{Frustrated crystallization
  of a monolayer of magnetized beads under geometrical confinement}},}\ }\href
  {\doibase 10.1103/physreve.95.062120} {\bibfield  {journal} {\bibinfo
  {journal} {Physical Review E}\ }\textbf {\bibinfo {volume} {95}},\ \bibinfo
  {pages} {062120} (\bibinfo {year} {2017})}\BibitemShut {NoStop}%
\bibitem [{\citenamefont {Opsomer}\ \emph {et~al.}(2020)\citenamefont
  {Opsomer}, \citenamefont {Merminod}, \citenamefont {Schockmel}, \citenamefont
  {Vandewalle}, \citenamefont {Berhanu},\ and\ \citenamefont
  {Falcon}}]{Opsomer:2020}%
  \BibitemOpen
  \bibfield  {author} {\bibinfo {author} {\bibfnamefont {E.}~\bibnamefont
  {Opsomer}}, \bibinfo {author} {\bibfnamefont {S.}~\bibnamefont {Merminod}},
  \bibinfo {author} {\bibfnamefont {J.}~\bibnamefont {Schockmel}}, \bibinfo
  {author} {\bibfnamefont {N.}~\bibnamefont {Vandewalle}}, \bibinfo {author}
  {\bibfnamefont {M.}~\bibnamefont {Berhanu}}, \ and\ \bibinfo {author}
  {\bibfnamefont {E.}~\bibnamefont {Falcon}},\ }\bibfield  {title} {\enquote
  {\bibinfo {title} {{Patterns in magnetic granular media at the crossover from
  two to three dimensions}},}\ }\href {\doibase 10.1103/physreve.102.042907}
  {\bibfield  {journal} {\bibinfo  {journal} {Physical Review E}\ }\textbf
  {\bibinfo {volume} {102}},\ \bibinfo {pages} {042907} (\bibinfo {year}
  {2020})}\BibitemShut {NoStop}%
\bibitem [{\citenamefont {Zahn}, \citenamefont {Lenke},\ and\ \citenamefont
  {Maret}(1998)}]{Zahn:1998}%
  \BibitemOpen
  \bibfield  {author} {\bibinfo {author} {\bibfnamefont {K.}~\bibnamefont
  {Zahn}}, \bibinfo {author} {\bibfnamefont {R.}~\bibnamefont {Lenke}}, \ and\
  \bibinfo {author} {\bibfnamefont {G.}~\bibnamefont {Maret}},\ }\bibfield
  {title} {\enquote {\bibinfo {title} {{Two-Stage Melting of Paramagnetic
  Colloidal Crystals in Two Dimensions}},}\ }\href {\doibase
  10.1103/physrevlett.82.2721} {\bibfield  {journal} {\bibinfo  {journal}
  {Physical Review Letters}\ }\textbf {\bibinfo {volume} {82}},\ \bibinfo
  {pages} {2721--2724} (\bibinfo {year} {1998})}\BibitemShut {NoStop}%
\bibitem [{\citenamefont {Gribova}\ \emph {et~al.}(2011)\citenamefont
  {Gribova}, \citenamefont {Arnold}, \citenamefont {Schilling},\ and\
  \citenamefont {Holm}}]{Gribova:2011}%
  \BibitemOpen
  \bibfield  {author} {\bibinfo {author} {\bibfnamefont {N.}~\bibnamefont
  {Gribova}}, \bibinfo {author} {\bibfnamefont {A.}~\bibnamefont {Arnold}},
  \bibinfo {author} {\bibfnamefont {T.}~\bibnamefont {Schilling}}, \ and\
  \bibinfo {author} {\bibfnamefont {C.}~\bibnamefont {Holm}},\ }\bibfield
  {title} {\enquote {\bibinfo {title} {{How close to two dimensions does a
  Lennard-Jones system need to be to produce a hexatic phase?}}}\ }\href
  {\doibase 10.1063/1.3623783} {\bibfield  {journal} {\bibinfo  {journal} {The
  Journal of Chemical Physics}\ }\textbf {\bibinfo {volume} {135}},\ \bibinfo
  {pages} {054514} (\bibinfo {year} {2011})},\ \Eprint
  {http://arxiv.org/abs/1104.0611} {1104.0611} \BibitemShut {NoStop}%
\bibitem [{\citenamefont {Komatsu}\ and\ \citenamefont
  {Tanaka}(2015)}]{Komatsu:2015}%
  \BibitemOpen
  \bibfield  {author} {\bibinfo {author} {\bibfnamefont {Y.}~\bibnamefont
  {Komatsu}}\ and\ \bibinfo {author} {\bibfnamefont {H.}~\bibnamefont
  {Tanaka}},\ }\bibfield  {title} {\enquote {\bibinfo {title} {{Roles of Energy
  Dissipation in a Liquid-Solid Transition of Out-of-Equilibrium Systems}},}\
  }\href {\doibase 10.1103/physrevx.5.031025} {\bibfield  {journal} {\bibinfo
  {journal} {Physical Review X}\ }\textbf {\bibinfo {volume} {5}},\ \bibinfo
  {pages} {031025} (\bibinfo {year} {2015})},\ \Eprint
  {http://arxiv.org/abs/1509.03435} {1509.03435} \BibitemShut {NoStop}%
\bibitem [{\citenamefont {Reis}, \citenamefont {Ingale},\ and\ \citenamefont
  {Shattuck}(2006)}]{Reis:2006}%
  \BibitemOpen
  \bibfield  {author} {\bibinfo {author} {\bibfnamefont {P.~M.}\ \bibnamefont
  {Reis}}, \bibinfo {author} {\bibfnamefont {R.~A.}\ \bibnamefont {Ingale}}, \
  and\ \bibinfo {author} {\bibfnamefont {M.~D.}\ \bibnamefont {Shattuck}},\
  }\bibfield  {title} {\enquote {\bibinfo {title} {{Crystallization of a
  Quasi-Two-Dimensional Granular Fluid}},}\ }\href {\doibase
  10.1103/physrevlett.96.258001} {\bibfield  {journal} {\bibinfo  {journal}
  {Physical Review Letters}\ }\textbf {\bibinfo {volume} {96}},\ \bibinfo
  {pages} {258001} (\bibinfo {year} {2006})},\ \Eprint
  {http://arxiv.org/abs/cond-mat/0603408} {cond-mat/0603408} \BibitemShut
  {NoStop}%
\bibitem [{\citenamefont {Behringer}\ and\ \citenamefont
  {Chakraborty}(2018)}]{Behringer:2018}%
  \BibitemOpen
  \bibfield  {author} {\bibinfo {author} {\bibfnamefont {R.~P.}\ \bibnamefont
  {Behringer}}\ and\ \bibinfo {author} {\bibfnamefont {B.}~\bibnamefont
  {Chakraborty}},\ }\bibfield  {title} {\enquote {\bibinfo {title} {The physics
  of jamming for granular materials: a review},}\ }\href {\doibase
  10.1088/1361-6633/aadc3c} {\bibfield  {journal} {\bibinfo  {journal} {Reports
  on Progress in Physics}\ }\textbf {\bibinfo {volume} {82}},\ \bibinfo {pages}
  {012601} (\bibinfo {year} {2018})}\BibitemShut {NoStop}%
\bibitem [{\citenamefont {Bi}\ \emph {et~al.}(2016)\citenamefont {Bi},
  \citenamefont {Yang}, \citenamefont {Marchetti},\ and\ \citenamefont
  {Manning}}]{Bi:2016}%
  \BibitemOpen
  \bibfield  {author} {\bibinfo {author} {\bibfnamefont {D.}~\bibnamefont
  {Bi}}, \bibinfo {author} {\bibfnamefont {X.}~\bibnamefont {Yang}}, \bibinfo
  {author} {\bibfnamefont {M.~C.}\ \bibnamefont {Marchetti}}, \ and\ \bibinfo
  {author} {\bibfnamefont {M.~L.}\ \bibnamefont {Manning}},\ }\bibfield
  {title} {\enquote {\bibinfo {title} {{Motility-Driven Glass and Jamming
  Transitions in Biological Tissues}},}\ }\href {\doibase
  10.1103/physrevx.6.021011} {\bibfield  {journal} {\bibinfo  {journal}
  {Physical Review X}\ }\textbf {\bibinfo {volume} {6}},\ \bibinfo {pages}
  {021011} (\bibinfo {year} {2016})},\ \Eprint
  {http://arxiv.org/abs/1509.06578} {1509.06578} \BibitemShut {NoStop}%
\bibitem [{\citenamefont {Lumay}\ \emph {et~al.}(2015)\citenamefont {Lumay},
  \citenamefont {Schockmel}, \citenamefont {Henández-Enríquez}, \citenamefont
  {Dorbolo}, \citenamefont {Vandewalle},\ and\ \citenamefont
  {Pacheco-Vázquez}}]{magnet}%
  \BibitemOpen
  \bibfield  {author} {\bibinfo {author} {\bibfnamefont {G.}~\bibnamefont
  {Lumay}}, \bibinfo {author} {\bibfnamefont {J.}~\bibnamefont {Schockmel}},
  \bibinfo {author} {\bibfnamefont {D.}~\bibnamefont {Henández-Enríquez}},
  \bibinfo {author} {\bibfnamefont {S.}~\bibnamefont {Dorbolo}}, \bibinfo
  {author} {\bibfnamefont {N.}~\bibnamefont {Vandewalle}}, \ and\ \bibinfo
  {author} {\bibfnamefont {F.}~\bibnamefont {Pacheco-Vázquez}},\ }\bibfield
  {title} {\enquote {\bibinfo {title} {Flow of magnetic repelling grains in a
  two-dimensional silo},}\ }\href@noop {} {\bibfield  {journal} {\bibinfo
  {journal} {Papers in Physics}\ }\textbf {\bibinfo {volume} {7}},\ \bibinfo
  {pages} {070013} (\bibinfo {year} {2015})}\BibitemShut {NoStop}%
\bibitem [{\citenamefont {Hernández-Enríquez}, \citenamefont {Lumay},\ and\
  \citenamefont {Pacheco-Vázquez}(2017)}]{Discharge}%
  \BibitemOpen
  \bibfield  {author} {\bibinfo {author} {\bibfnamefont {D.}~\bibnamefont
  {Hernández-Enríquez}}, \bibinfo {author} {\bibfnamefont {G.}~\bibnamefont
  {Lumay}}, \ and\ \bibinfo {author} {\bibfnamefont {F.}~\bibnamefont
  {Pacheco-Vázquez}},\ }\bibfield  {title} {\enquote {\bibinfo {title}
  {Discharge of repulsive grains from a silo: experiments and simulations},}\
  }\href@noop {} {\bibfield  {journal} {\bibinfo  {journal} {EPJ Web Conf.}\
  }\textbf {\bibinfo {volume} {140}},\ \bibinfo {pages} {03089} (\bibinfo
  {year} {2017})}\BibitemShut {NoStop}%
\bibitem [{\citenamefont {Escobar-Ortega}\ \emph {et~al.}(2020)\citenamefont
  {Escobar-Ortega}, \citenamefont {Hidalco-Caballero}, \citenamefont
  {Marston},\ and\ \citenamefont {Pacheco-Vázquez}}]{viscoelastic}%
  \BibitemOpen
  \bibfield  {author} {\bibinfo {author} {\bibfnamefont {Y.~Y.}\ \bibnamefont
  {Escobar-Ortega}}, \bibinfo {author} {\bibfnamefont {S.}~\bibnamefont
  {Hidalco-Caballero}}, \bibinfo {author} {\bibfnamefont {J.~O.}\ \bibnamefont
  {Marston}}, \ and\ \bibinfo {author} {\bibfnamefont {F.}~\bibnamefont
  {Pacheco-Vázquez}},\ }\bibfield  {title} {\enquote {\bibinfo {title} {The
  viscoelastic-like response of a repulsive granular medium during projectile
  impact and penetration},}\ }\href@noop {} {\bibfield  {journal} {\bibinfo
  {journal} {Journal of Non-Newtonian Fluid Mechanics}\ }\textbf {\bibinfo
  {volume} {280}},\ \bibinfo {pages} {104295} (\bibinfo {year}
  {2020})}\BibitemShut {NoStop}%
\bibitem [{\citenamefont {Modesto}\ \emph {et~al.}(2022)\citenamefont
  {Modesto}, \citenamefont {Dorbolo}, \citenamefont {Katsuragi}, \citenamefont
  {Pacheco-Vázquez},\ and\ \citenamefont {Sobral}}]{compression}%
  \BibitemOpen
  \bibfield  {author} {\bibinfo {author} {\bibfnamefont {J.~A.~C.}\
  \bibnamefont {Modesto}}, \bibinfo {author} {\bibfnamefont {S.}~\bibnamefont
  {Dorbolo}}, \bibinfo {author} {\bibfnamefont {H.}~\bibnamefont {Katsuragi}},
  \bibinfo {author} {\bibfnamefont {F.}~\bibnamefont {Pacheco-Vázquez}}, \
  and\ \bibinfo {author} {\bibfnamefont {Y.~D.}\ \bibnamefont {Sobral}},\
  }\bibfield  {title} {\enquote {\bibinfo {title} {Experimental and numerical
  investigation of the compression and expansion of a granular bed of repelling
  magnetic disks},}\ }\href@noop {} {\bibfield  {journal} {\bibinfo  {journal}
  {Granular Matter}\ }\textbf {\bibinfo {volume} {24}},\ \bibinfo {pages} {105}
  (\bibinfo {year} {2022})}\BibitemShut {NoStop}%
\bibitem [{\citenamefont {Yu}\ \emph {et~al.}(2017)\citenamefont {Yu},
  \citenamefont {Ahmad}, \citenamefont {Ståhl}, \citenamefont {Su},
  \citenamefont {Glazyrin}, \citenamefont {Liermann}, \citenamefont {Franz},
  \citenamefont {Cao}, \citenamefont {Zhang},\ and\ \citenamefont
  {Jiang}}]{alloy}%
  \BibitemOpen
  \bibfield  {author} {\bibinfo {author} {\bibfnamefont {Q.}~\bibnamefont
  {Yu}}, \bibinfo {author} {\bibfnamefont {A.~S.}\ \bibnamefont {Ahmad}},
  \bibinfo {author} {\bibfnamefont {K.}~\bibnamefont {Ståhl}}, \bibinfo
  {author} {\bibfnamefont {Y.}~\bibnamefont {Su}}, \bibinfo {author}
  {\bibfnamefont {K.}~\bibnamefont {Glazyrin}}, \bibinfo {author}
  {\bibfnamefont {H.~P.}\ \bibnamefont {Liermann}}, \bibinfo {author}
  {\bibfnamefont {H.}~\bibnamefont {Franz}}, \bibinfo {author} {\bibfnamefont
  {Q.~P.}\ \bibnamefont {Cao}}, \bibinfo {author} {\bibfnamefont {D.~X.}\
  \bibnamefont {Zhang}}, \ and\ \bibinfo {author} {\bibfnamefont {J.~Z.}\
  \bibnamefont {Jiang}},\ }\bibfield  {title} {\enquote {\bibinfo {title}
  {Pressure-induced structural change in liquid gain eutectic alloy},}\
  }\href@noop {} {\bibfield  {journal} {\bibinfo  {journal} {Scientific
  Reports}\ }\textbf {\bibinfo {volume} {7}},\ \bibinfo {pages} {1139}
  (\bibinfo {year} {2017})}\BibitemShut {NoStop}%
\bibitem [{\citenamefont {Finney}(1977)}]{Finney:1977}%
  \BibitemOpen
  \bibfield  {author} {\bibinfo {author} {\bibfnamefont {J.~L.}\ \bibnamefont
  {Finney}},\ }\bibfield  {title} {\enquote {\bibinfo {title} {{Modelling the
  structures of amorphous metals and alloys}},}\ }\href {\doibase
  10.1038/266309a0} {\bibfield  {journal} {\bibinfo  {journal} {Nature}\
  }\textbf {\bibinfo {volume} {266}},\ \bibinfo {pages} {309--314} (\bibinfo
  {year} {1977})}\BibitemShut {NoStop}%
\bibitem [{\citenamefont {Finney}\ and\ \citenamefont
  {Bernal}(1970)}]{Finney:1970}%
  \BibitemOpen
  \bibfield  {author} {\bibinfo {author} {\bibfnamefont {J.~L.}\ \bibnamefont
  {Finney}}\ and\ \bibinfo {author} {\bibfnamefont {J.~D.}\ \bibnamefont
  {Bernal}},\ }\bibfield  {title} {\enquote {\bibinfo {title} {{Random packings
  and the structure of simple liquids. I. The geometry of random close
  packing}},}\ }\href {\doibase 10.1098/rspa.1970.0189} {\bibfield  {journal}
  {\bibinfo  {journal} {Proceedings of the Royal Society of London. A.
  Mathematical and Physical Sciences}\ }\textbf {\bibinfo {volume} {319}},\
  \bibinfo {pages} {479--493} (\bibinfo {year} {1970})},\ \bibinfo {note} {doi:
  10.1098/rspa.1970.0189}\BibitemShut {NoStop}%
\end{thebibliography}%

\end{document}